\documentstyle[preprint,aps]{revtex}
\begin{document}
\draft

\title
{Dirac's Observables for the Higgs Model: I) the Abelian Case.}

\author{Luca Lusanna}

\address
{Sezione INFN di Firenze\\
L.go E.Fermi 2 (Arcetri)\\
50125 Firenze, Italy\\
E-mail LUSANNA@FI.INFN.IT}

\author{and}

\author{Paolo Valtancoli}

\address
{Dipartimento di Fisica\\
Universita' di Firenze\\
L.go E.Fermi 2 (Arcetri)\\
50125 Firenze, Italy\\
E-mail VALTANCOLI@FI.INFN.IT}

\maketitle

\begin{abstract}

We search a canonical basis of Dirac's observables for the classical
Abelian Higgs model with fermions in the case of a trivial U(1) principal 
bundle. The study of the Gauss law first class constraint shows that the
model has two disjoint sectors of solutions associated with two physically 
different phases. In the electromagnetic phase,
the electromagnetic field remains massless: after the determination of the
Dirac's observables we get that both the reduced physical Hamiltonian and
Lagrangian are nonlocal. In the Higgs phase, the electromagnetic field
becomes massive and in terms of Dirac's observables we get a local, but
nonanalytic in the electric charge (or equivalently in the sum of the 
electromagnetic mass and of the residual Higgs field), 
physical Hamiltonian; however the associated Lagrangian is nonlocal.
Some comments on the R-gauge-fixing, the possible elimination of the
residual Higgs field and on the Nielsen-Olesen vortex solution close the
paper.

\vskip 1truecm
\noindent February 1996
\vskip 1truecm
\noindent This work has been partially supported by the network ``Constrained 
Dynamical Systems" of the E.U. Programme ``Human Capital and Mobility".

\end{abstract}
\vfill\eject

\section
{Introduction}

After having found a symplectic basis of Dirac's observables for the
classical Yang-Mills theory with Grassmann-valued fermions Ref.\cite{lus}
in the case of a trivial principal bundle over Minkowski spacetime and in
suitable function spaces where the Gribov ambiguity is absent, the next step in
the program \cite{lus1}
of reformulating particle physics in terms of Dirac's observables
is the study of the Higgs model. This model is needed to generate the
spontaneous symmetry breaking used in the $SU(2)\times U(1)$ electroweak
standard model to give mass to the vector gauge bosons and, through the
Yukawa couplings, to the fermions. Here, we shall preliminary study the
classical Abelian Higgs model with fermions [trivial U(1) principal bundle]
to disentangle the basic implications of the Higgs mechanism from the
complications of the $SU(2)\times U(1)$ model.

The Abelian Higgs model is described by the following Lagrangian density
[$\lambda > 0$, $\phi_o > 0$]

\begin{eqnarray}
{\cal L}(x)&=&-{1\over 4}F_{\mu\nu}(x)F^{\mu\nu}(x)+[D^{(A)}_{\mu}\phi (x)]
^{*}\, D^{(A)\mu} \phi (x) -V(\phi )+
\nonumber \\
&&+{1\over2}\bar \psi (x)[\gamma^{\mu}(i\partial_{\mu}+eA_{\mu}(x))
-\stackrel{\longleftarrow}{(i\partial_{\mu}-eA_{\mu}(x))}\gamma^{\mu}]\psi (x)
- m\bar \psi (x)\psi (x)\nonumber \\
V(\phi )&=&\lambda {[\phi^{*}(x)\phi (x)-\phi^2_o]}^2=\mu^2\phi^{*}(x)\phi
(x)+\lambda [\phi^{*}(x)\phi (x)]^2+\lambda \phi_o^4=\nonumber \\
&=&-{1\over 2}m^2_H\phi^{*}(x)\phi (x)+\lambda [\phi^{*}(x)\phi (x)]^2+
\lambda \phi_o^4,\nonumber \\
&&{}\nonumber \\
&&\mu^2=-2\lambda \phi_o^2 < 0,
\quad\quad m^2_H=-2\mu^2=4\lambda \phi_o^2,\quad\quad
\phi_o={{m_H}\over {2\sqrt{\lambda}}}=\sqrt{ {{-\mu^2}\over {2\lambda}} },
\label{1}
\end{eqnarray}

\noindent where $\phi (x)$, the Higgs field, is a complex scalar field
[$D^{(A)}_{\mu}\phi (x)=(\partial_{\mu}-ieA_{\mu}(x))\phi (x)$], 
$\mu^2 < 0$ so that the potential $V(\phi )$ has a set of absolute
minima for $\phi^{*}\phi =\phi_o^2$, parametrized by a phase [$\phi
\mapsto e^{i\theta}\phi$ leaves $\phi^{*}\phi$ invariant], and with
$\phi_o > 0$ an arbitrary real number [$< \phi > = \phi_o\not= 0$
at the quantum level: this is the gauge non-invariant formulation
of the statement of symmetry breaking].The fermion field
$\psi (x)$ is Grassmann-valued and
is absent when the Abelian Higgs model is used as a relativistic
generalization \cite{eb,sho} of the Landau-Ginzburg treatment \cite{gl}
of superconductivity with $\phi (x)$ [also called the ``complex order
parameter", with the ordered phase being the broken symmetry one]
associated with the spin-singlet part of the 
nonzero vacuum expectation value of the fermion bilinear describing the
Bose-Einstein condensation of the Cooper electron pairs [the attractive effects 
of virtual phonons being slightly higher of the Coulomb repulsion] and with
the massless Goldstone boson, generated by the spontaneous symmetry
breaking, reabsorbed to 
give mass to the photon so to obtain a finite-range electromagnetic field
as required by the Meissner effect of magnetic flux exclusion \cite{an}
(physically the longitudinal degrees of freedom of the electromagnetic field 
couple to the plasma oscillations, i.e. to the collective density fluctuations 
of the electrons).

The Lagrangian density is invariant under the U(1) gauge transformations
$A_{\mu}(x)\mapsto A_{\mu}(x)-{1\over e}\partial_{\mu}\alpha (x)$, $\phi (x)
\mapsto e^{-i\alpha (x)}\phi (x)$, $\psi (x)\mapsto e^{-i\alpha (x)}\psi
(x)$.

We shall show that the singular Lagrangian density of Eq.(\ref{1}) describes
simultaneously two extremely different dynamics, since its associate Gauss
law constraint (or equivalently the corresponding acceleration-independent
Euler-Lagrange equation) generates two disjoint sectors of solutions and only
one of them (the electromagnetic phase with massless electromagnetic fields)
is analytic in the coupling constant (the electric charge). To describe these 
two sectors, i.e. the electromagnetic and Higgs phases respectively, we shall 
use different parametrizations of the Higgs fields.

\section
{The Electromagnetic Phase}

The canonical momenta associated with Eq.(\ref{1}) are

\begin{eqnarray}
&&\pi^o(x)=0\nonumber \\
&&\vec \pi (x)=\vec E (x)\nonumber \\
&&\pi (x)=-{i\over 2}\psi^{\dagger}(x)\nonumber \\
&&\bar \pi (x)=-{i\over 2}\gamma^o\psi (x)\nonumber \\
&&\pi_{\phi}(x)={\dot \phi}^{*}(x)+ieA_o(x)\phi^{*}(x)\nonumber \\
&&\pi_{\phi {*}}(x)=\dot \phi (x)-ieA_o(x)\phi (x),
\label{2}
\end{eqnarray}

\noindent and are assumed to satisfy the Poisson brackets

\begin{eqnarray}
&&\lbrace A_{\mu}(\vec x,x^o), \pi^{\nu}(\vec y,x^o)\rbrace =\delta^{\nu}_{\mu}
\delta^3(\vec x-\vec y)\nonumber \\
&&\lbrace \psi_{\alpha}(\vec x,x^o),\pi_{\beta}(\vec y,x^o)\rbrace =
\lbrace {\bar \psi}_{\alpha}(\vec x,x^o),{\bar \pi}_{\beta}(\vec y,x^o)
\rbrace =-\delta_{\alpha\beta}\delta^3(\vec x-\vec y)\nonumber \\
&&\lbrace \phi (\vec x,x^o),\pi_{\phi}(\vec y,x^o)\rbrace =
\lbrace \phi^{*}(\vec x,x^o), \pi_{\phi {*}}(\vec y,x^o)\rbrace =
\delta^3(\vec x-\vec y).
\label{3}
\end{eqnarray}

\noindent By eliminating the fermionic second class constraints with the
introduction of the Dirac brackets

\begin{equation}
\lbrace \psi_{\alpha}(\vec x,x^o),{\bar \psi}_{\beta}(\vec y,x^o)\rbrace
{}^{*}=-i{(\gamma^o)}_{\alpha\beta}\delta^3(\vec x-\vec y)
\label{4}
\end{equation}

\noindent (denoted $\lbrace .,.\rbrace$ in the rest of the paper for the
sake of simplicity) as shown in Ref.\cite{lus}, one arrives at the following 
Dirac Hamiltonian density [$\lambda_o(x)$ is a Dirac multiplier and an
integration by parts has been done]

\begin{eqnarray}
{\cal H}_D(x)&=&{1\over 2}[{\vec \pi}^2(x)+{\vec B}^2(x)]+\psi^{\dagger}(x)
\vec \alpha \cdot (i\vec \partial +e\vec A (x))\psi (x) +m\bar \psi (x)
\psi (x)+\nonumber \\
&+&\pi_{\phi {*}}(x)\pi_{\phi}(x)+[(\vec \partial +ie\vec A(x))\phi^{*}(x)]
\cdot (\vec \partial -ie\vec A(x))\phi (x)+\lambda {(\phi^{*}(x)\phi (x)-
\phi_o^2)}^2-\nonumber \\
&-&A_o(x)[-\vec \partial \cdot \vec \pi (x)+e\psi^{\dagger}(x)\psi (x)-ie
(\pi_{\phi}(x)\phi (x)-\pi_{\phi {*}}(x)\phi^{*}(x))]+\lambda_o(x)
\pi^o(x).
\label{5}
\end{eqnarray}

The constraint analysis shows that there a primary first class constraint,
$\pi^o(x)\approx 0$, and a secondary first class one (the Gauss law, namely
the acceleration-independent Euler-Lagrange equation of the model)

\begin{equation}
\Gamma (x) =-\vec \partial \cdot \vec \pi (x)+e\psi^{\dagger}(x)\psi (x)-
ie[\pi_{\phi}(x)\phi (x)-\pi_{\phi {*}}(x)\phi^{*}(x)]\approx 0.
\label{6}
\end{equation}

\noindent Eq.(\ref{6}) is ambiguous, since it can be considered either as an
elliptic equation for the electric field $\vec \pi$ or as an algebraic
equation in the Higgs momenta: in the first case one obtains a sector of 
solutions corresponding to the electromagnetic phase, in the second one the 
sector of the Higgs phase. As a consequence, the space of solutions of the
Euler-Lagrange equations is not connected being formed by two disjoint
subspaces (its zeroth homotopy group is not trivial).

Since the conserved energy-momentum and angular momentum tensor densities 
and Poincar\'e generators are ["${\buildrel \circ \over =}$" means
evaluated on the extremals of the action $S=\int d^4x {\cal L}(x)$; $\sigma
^{\mu\nu}={i\over 2}[\gamma^{\mu},\gamma^{\nu}]$, $\sigma^i={1\over 2}
\epsilon^{ijk}\sigma^{jk}$, $\vec \alpha =\gamma^o\vec \gamma$, $\beta =
\gamma^o$]

\begin{eqnarray}
\Theta^{\mu\nu}(x)&=&F^{\mu\alpha}(x)F_{\alpha}{}^{\nu}(x)+{1\over 4}\eta
^{\mu\nu}F^{\alpha\beta}(x)F_{\alpha\beta}(x)+\nonumber \\
&+&{1\over 2}[\bar \psi (x)\gamma
^{\mu}(i\partial^{\nu}+eA^{\nu}(x))\psi (x)-\bar \psi (x)
\stackrel{\longleftarrow}{(i\partial^{\nu}-eA^{\nu}(x))}\gamma^{\mu}\psi (x)]+
\nonumber \\
&+&{(D^{(A)\mu}\phi (x))}^{*}\, D^{(A)\nu}\phi (x)+{(D^{(A)\nu}\phi (x))}^{*}
D^{(A)\mu}\phi (x)-\nonumber \\
&-&\eta^{\mu\nu}[{(D^{(A)\alpha}\phi (x))}^{*}D^{(A)}{}_{\alpha}
\phi (x)-V(\phi )],\nonumber \\
{\cal M}^{\mu\alpha\beta}(x)&=&x^{\alpha}\Theta^{\mu\beta}(x)-x^{\beta}
\Theta^{\mu \alpha}(x)+{1\over 4}\bar \psi (x)(\gamma^{\mu}\sigma^{\alpha\beta}
+\sigma^{\alpha\beta}\gamma^{\mu})\psi (x),\nonumber \\
&&\partial_{\nu}\Theta^{\nu\mu}(x){\buildrel \circ \over =} 0,\quad\quad
\partial_{\mu}{\cal M}^{\mu\alpha\beta}(x){\buildrel \circ \over =} 0,
\nonumber \\
&&{}\nonumber \\
P^{\mu}&=&\int d^3x \Theta^{o\mu}(\vec x,x^o),\nonumber \\
J^{\mu\nu}&=&\int d^3x {\cal M}^{o\mu\nu}(\vec x,x^o),\nonumber \\
&&{}\nonumber \\
P^o&=&\int d^3x\, \lbrace {1\over 2}[{\vec \pi}^2(\vec x,x^o)+{\vec B}^2(\vec x,
x^o)]+\nonumber \\
+&\pi_{\phi}&(\vec x,x^o)\pi_{\phi^{*}}(\vec x,x^o)+{({\vec D}^{(A)}\phi
(\vec x,x^o))}^{*}\cdot {\vec D}^{(A)}\phi (\vec x,x^o)+V(\phi )+\nonumber \\
+&{1\over 2}&[\bar \psi (\vec x,x^o)\gamma^o(i\partial^o+eA^o(\vec x,x^o))\psi
(\vec x,x^o)-\bar \psi (\vec x,x^o)\stackrel{\longleftarrow}{(i\partial^o-
eA^o(\vec x,x^o))}\gamma^o\psi (\vec x,x^o)]\rbrace \nonumber \\
P^i&=&\int d^3x\, \lbrace {(\vec \pi (\vec x,x^o)\times \vec B(\vec x,x^o))}^i+
\nonumber \\
+&\pi_{\phi}&(\vec x,x^o)D^{(A)i}\phi (\vec x,x^o)+{(D^{(A)i}\phi
(\vec x,x^o))}^{*}\pi_{\phi^{*}}(\vec x,x^o)+\nonumber \\
+&{1\over 2}&[\bar \psi (\vec x,x^o)\gamma^o(i\partial^i+eA^i(\vec x,x^o))\psi
(\vec x,x^o)-\bar \psi (\vec x,x^o)\stackrel{\longleftarrow}{(i\partial^i-
eA^i(\vec x,x^o))}\gamma^o\psi (\vec x,x^o)]\rbrace \nonumber \\
J^i&=&{1\over 2}\epsilon^{ijk}J^{jk}=\int d^3x\, \lbrace {[\vec x\times({\vec
\pi}(\vec x,x^o)\times \vec B(\vec x,x^o))]}^i-\nonumber \\
-&&{[\vec x\times (\pi_{\phi}(\vec x,x^o){\vec D}^{(A)}\phi (\vec x,x^o)+
({\vec D}^{(A)}\phi (\vec x,x^o))^{*}\pi_{\phi^{*}}(\vec x,x^o))]}^i+
\nonumber \\
&+&{1\over 2}[\vec x\times [\bar \psi (\vec x,x^o)\gamma^o(i\vec \partial+
e\vec A(\vec x,x^o))\psi(\vec x,x^o)-\nonumber \\
&-&\bar \psi (\vec x,x^o)\stackrel{\longleftarrow}{(i\vec \partial-
e\vec A(\vec x,x^o))}\gamma^o\psi (\vec x,x^o)]\, ]^i\rbrace \nonumber \\
&+&{1\over 2}\psi^{\dagger} (\vec x,x^o)\sigma^i
\psi (\vec x,x^o)\rbrace \nonumber \\
K^i&=&J^{oi}=x^oP^i-\int d^3x\, x^i \Theta^{oo}(\vec x,x^o),
\label{7}
\end{eqnarray}

\noindent
following Dirac\cite{dir} and Ref.\cite{lus}, we will
assume boundary conditions $A_o(\vec x,x^o){\rightarrow}_{r\rightarrow
\infty}\, a_o/r^{1+\epsilon}$, $\vec A(\vec x,x^o){\rightarrow}_{r\rightarrow
\infty}\vec a/r^{2+\epsilon}$, $r=\, |\, \vec x\, |$, so that the Laplacian
on $R^3$, $\triangle =-{\vec \partial}^2$, has no zero modes and the
Poincar\'e generators are finite; this requires also the following
boundary conditions on the fermion and Higgs fields: $\psi(\vec x,x^o)
{\rightarrow}_{r\rightarrow \infty}\, \chi/r^{3/2+\epsilon}+O(r^{-2})$,
$\phi (\vec x,x^o){\rightarrow}_{r\rightarrow \infty}\, const.+\varphi /
r^{2+\epsilon}+O(r^{-3})$ [the ``constant" is required for the Higgs sector],
$\pi_{\phi}(\vec x,x^o){\rightarrow}_{r\rightarrow 
\infty}\, \zeta /r^{2+\epsilon}+O(r^{-3})$. The U(1) gauge transformations
are assumed to behave as $U(\vec x,x^o){\rightarrow}_{r\rightarrow \infty}\,
const. + O(r^{-1})$, so to preserve the boundary conditions. The previous 
boundary conditions are adapted to the fixed $x^o$, not Lorentz-covariant, 
Hamiltonian formalism; however, they are natural in its covariantization
by means of the reformulation of the theory on spacelike hypersurfaces
(see Section V).

In the electromagnetic phase one obtains
the following decompositions from the Hodge theorem

\begin{eqnarray}
\vec A(x)&=& \vec \partial \eta (x)+{\vec A}_{\perp}(x)\nonumber \\
\vec \pi (x)&=&{\vec \pi}_{\perp}(x)+{{\vec \partial}\over {\triangle}}
\lbrace \Gamma (x)-e\psi^{\dagger}(x)\psi (x)+ie[\pi_{\phi}(x)\phi (x)-
\pi_{\phi {*}}(x)\phi^{*}(x)]\rbrace \nonumber \\
&&{}\nonumber \\
&&\eta (\vec x,x^o)=-{1\over {\triangle}}\vec \partial \cdot \vec A(\vec x,x^o)
=-\int d^3y\, \vec c(\vec x-\vec y)\cdot \vec A(\vec y,x^o)\nonumber \\
&&\vec c(\vec x)={{\vec \partial}\over {\triangle}}\delta^3(\vec x-\vec y)
={{\vec x}\over {4\pi {|\, \vec x\, |}^3}}\nonumber \\
&&A^i_{\perp}(x)=(\delta^{ij}+{{\partial^i\partial^j}\over {\triangle}})
A^j(x),\quad \pi^i_{\perp}(x)=(\delta^{ij}+{{\partial^i\partial^j}\over 
{\triangle}})\pi^j(x)\nonumber \\
&&{}\nonumber \\
&&\lbrace \eta (\vec x,x^o),\Gamma (\vec y,x^o)\rbrace =-\delta^3(\vec x-\vec y)
\nonumber \\
&&\lbrace A^i_{\perp}(\vec x,x^o),\pi^j_{\perp}(\vec y,x^o)\rbrace =-
(\delta^{ij}+{{\partial^i\partial^j}\over {\triangle}})\delta^3(\vec x-\vec y).
\label{8}
\end{eqnarray}

\noindent The fields $A_o(x), \pi^o(x)$ and $\eta (x), \Gamma (x)$ are pairs of
conjugate gauge variables, while ${\vec A}_{\perp}(x), {\vec \pi}_{\perp}
(x)$ are a canonical basis of Dirac's observables. As shown in Ref.\cite{lus},
the Dirac observables for the fermion field are

\begin{eqnarray}
\check \psi (x)&=& e^{-ie\eta (x)}\psi (x)\nonumber \\
{\check \psi}^{\dagger}(x)&=& \psi^{\dagger}(x) e^{ie\eta (x)}\nonumber \\
&&\lbrace {\check \psi}_{\alpha}(\vec x,x^o),{\check \psi}^{\dagger}_{\beta}
(\vec y,x^o)\rbrace =-i\delta_{\alpha\beta}\delta^3(\vec x-\vec y);
\label{9}
\end{eqnarray}

\noindent they describe the charged fermions dressed with their Coulomb cloud.

Since

\begin{eqnarray}
&&\lbrace \phi (\vec x,x^o),\Gamma (\vec y,x^o)\rbrace =-ie\phi (\vec x,x^o)
\delta^3(\vec x-\vec y),\nonumber \\
&&\lbrace \pi_{\phi}(\vec x,x^o),\Gamma (\vec y,x^o)\rbrace =+ie\pi_{\phi}
(\vec x,x^o)\delta^3(\vec x-\vec y),
\label{10}
\end{eqnarray}

\noindent the Dirac observables for the Higgs field are

\begin{eqnarray}
\check \phi (x)&=& e^{-ie\eta (x)}\phi (x),\nonumber \\
{\check \pi}_{\phi}(x)&=& e^{ie\eta (x)}\pi_{\phi}(x),\nonumber \\
&&\lbrace \check \phi (\vec x,x^o),\Gamma (\vec y,x^o)\rbrace = \lbrace
{\check \pi}_{\phi}(\vec x,x^o),\Gamma (\vec y,x^o)\rbrace =0,
\label {11}
\end{eqnarray}

\noindent and again it amounts to add the Coulomb cloud to them. 

Therefore, the physical Hamiltonian density after the symplectic decoupling 
of the gauge variables is

\begin{eqnarray}
{\cal H}^{(em)}_{phys}(x)&=&{1\over 2}[{\vec \pi}^2_{\perp}(x)+{\vec B}^2
(x)]+{\check \psi}^{\dagger}(x)\vec \alpha \cdot (i\vec \partial +e{\vec A}
_{\perp}(x))\check \psi (x)+m{\check {\bar \psi}}(x)\check \psi (x)+
\nonumber \\
&+&{\check \pi}_{\phi}(x){\check \pi}_{\phi {*}}(x)+[(\vec \partial +
ie{\vec A}_{\perp}(x)){\check \phi}^{*}(x)]\cdot (\vec \partial -
ie{\vec A}_{\perp}(x))\check \phi (x)+\nonumber \\
&+&\lambda {({\check \phi}^{*}(x)\check \phi (x)-\phi_o^2)}^2+\nonumber \\
&+&{{e^2}\over 2}[{\check \psi}^{\dagger}(x)\check \psi (x)-
i({\check \pi}_{\phi}(x)\check \phi (x)-{\check \pi}_{\phi {*}}(x){\check \phi}
^{*}(x))]\, {1\over {\triangle}}\nonumber \\
&&[{\check \psi}^{\dagger}(x)\check \psi (x)-
i({\check \pi}_{\phi}(x)\check \phi (x)-{\check \pi}_{\phi {*}}(x){\check \phi}
^{*}(x))].
\label{12}
\end{eqnarray}

This Hamiltonian density is analytic in the electric charge e but there is the
nonlocal Coulomb interaction of the charged fields $\check \psi ,{\check
\psi}^{\dagger}, \check \phi ,{\check \phi}^{*}$. 
See Refs.\cite{lus,lus2} and Section V for the
reformulation on spacelike hypersurfaces to take care of Lorentz covariance.

The Hamilton equations imply

\begin{eqnarray}
{\vec \pi}_{\perp}(x)&=&-\partial^o {\vec A}_{\perp}(x),\nonumber \\
{\check \pi}_{\phi}(x)&=&\partial^o {\check \phi}^{*}(x)-ie^2{\check \phi}^{*}
(x){1\over {\triangle}}[{\check \psi}^{\dagger}(x){\check \psi}(x)+i
({\check \pi}_{\phi^{*}}(x){\check \phi}^{*}(x)-{\check \pi}_{\phi}(x){\check
\phi}(x))]=\nonumber \\
&=&\partial^o {\check \phi}^{*}(x)-\nonumber \\
&-&ie^2{\check \phi}^{*}(x){1\over {\triangle +
2e^2{\check \phi}^{*}(x){\check \phi}(x)}}[{\check \psi}^{\dagger}(x){\check 
\psi}(x)+i({\check \phi}^{*}(x)\partial {\check \phi}(x)-\partial^o 
{\check \phi}^{*}(x){\check \phi}(x))]\nonumber \\
{\check \pi}_{\phi {*}}(x)&=&\partial^o {\check \phi}(x)+ie^2{\check \phi}(x)
{1\over {\triangle}}[{\check \psi}^{\dagger}(x){\check \psi}(x)+i
({\check \pi}_{\phi^{*}}(x){\check \phi}^{*}(x)-{\check \pi}_{\phi}(x){\check
\phi}(x))]=\nonumber \\
&=&\partial^o {\check \phi}(x)+\nonumber \\
&+&ie^2{\check \phi}(x){1\over {\triangle +
2e^2{\check \phi}^{*}(x){\check \phi}(x)}}[{\check \psi}^{\dagger}(x){\check 
\psi}(x)+i({\check \phi}^{*}(x)\partial^o {\check \phi}(x)-\partial^o 
{\check \phi}^{*}(x){\check \phi}(x))]
\label{13}
\end{eqnarray}

\noindent because we get

\begin{eqnarray}
&&{\check \pi}_{\phi^{*}}(x){\check \phi}^{*}(x)-{\check \pi}_{\phi}(x){\check
\phi}(x)=\nonumber \\
&&{1\over {1+2e^2{\check \phi}^{*}(x){\check \phi}(x){1\over {\triangle}} }}
[\partial^o {\check \phi}(x){\check \phi}^{*}(x)-
\partial^o {\check \phi}^{*}(x)
{\check \phi}(x)+2ie^2{\check \phi}^{*}(x){\check \phi}(x){1\over {\triangle}}
{\check \psi}^{\dagger}(x){\check \psi}(x)]
\label{14}
\end{eqnarray}

\noindent Use has been done of the operator identity ${1\over A}{1\over
{1+B{1\over A}}}={1\over A}[1-B{1\over A}+B{1\over A}B{1\over A}-...]=
{1\over {A+B}}$ (valid for B a small perturbation of  A) for $A=
\triangle$ and $B=\phi^{*}(x)\phi (x)$.

The nonlocal Lagrangian density generating ${\cal H}^{(em)}_{phys}(x)$ and
describing only the electromagnetic phase is (see also Ref.\cite{lus})

\begin{eqnarray}
{\cal L}^{(em)}_{phys}(x)&=&{1\over 2}[{\dot {\vec A}}^2_{\perp}(x)-{(\vec 
\partial \times {\vec A}_{\perp}(x))}^2]+\nonumber \\
&+&{\check \psi}^{\dagger}(x)[i\partial_o-\vec \alpha \cdot (i\vec \partial +
e{\vec A}_{\perp}(x))-\beta m]\check \psi (x)+\nonumber \\
&+&\partial^o {\check \phi}^{*}(x)\partial^o {\check \phi}(x)-[(\vec \partial +
ie{\vec A}_{\perp}(x)){\check \phi}^{*}(x)]\cdot (\vec \partial -ie{\vec A}
_{\perp}(x))\check \phi (x)-\nonumber \\
&-&\lambda {({\check \phi}^{*}(x)\check \phi (x)-\phi_o^2)}^2-\nonumber \\
&-&{{e^2}\over 2}[{\check \psi}^{\dagger}(x)\check \psi (x)+
i({\check \phi}^{*}(x)\partial^o{\check \phi}(x)-
\partial^o{\check \phi}^{*}(x){\check \phi}(x))]\, 
{1\over {\triangle +2e^2{\check \phi}^{*}(x){\check \phi}(x)}}\cdot \nonumber \\
&&[{\check \psi}^{\dagger}(x)\check \psi (x)+
i({\check \phi}^{*}(x)\partial^o{\check \phi}(x)-
\partial^o{\check \phi}^{*}(x){\check \phi}(x))],
\label{15}
\end{eqnarray}

The Hamilton equations of this phase are Eqs.(\ref{13}) and

\begin{eqnarray}
&&\partial^o{\vec \pi}_{\perp}(\vec x,x^o){\buildrel \circ \over =}
\triangle {\vec A}_{\perp}(\vec x,x^o)+e{\check \psi}^{\dagger}(\vec x,x^o)\vec
\alpha \check \psi (\vec x,x^o)+\nonumber \\
&+&ie[{\check \phi}^{*}(\vec x,x^o)(\vec \partial -ie{\vec A}_{\perp}
(\vec x,x^o))\check \phi (\vec x,x^o)-\check \phi (\vec x,x^o)
(\vec \partial +ie{\vec A}_{\perp}(\vec x,x^o)){\check \phi}^{*}
(\vec x,x^o)\nonumber \\
&&{}\nonumber \\
&&\partial^o{\check \pi}_{\phi}(\vec x,x^o){\buildrel \circ \over =}
(\vec \partial +ie{\vec A}_{\perp}(\vec x,x^o))^2{\check \phi}^{*}(\vec x,x^o)
-2\lambda {\check \phi}^{*}(\vec x,x^o)
[{\check \phi}^{*}(\vec x,x^o)\check \phi (\vec x,x^o)-\phi_o^2]+
\nonumber \\
&+&ie^2{\check \pi}_{\phi}(\vec x,x^o){1\over {\triangle}}[{\check \psi}
^{\dagger}(\vec x,x^o)\check \psi (\vec x,x^o)-i({\check \pi}_{\phi}(\vec x,x^o)
\check \phi (\vec x,x^o)-{\check \phi}_{\phi^{*}}(\vec x,x^o){\check \phi}
^{*}(\vec x,x^o))]\nonumber \\
&&{}\nonumber \\
&&\partial^o{\check \pi}_{\phi^{*}}(\vec x,x^o){\buildrel \circ \over =}
(\vec \partial +ie{\vec A}_{\perp}(\vec x,x^o))^2{\check \phi}^{*}(\vec x,x^o)-
2\lambda \check \phi (\vec x,x^o)
[{\check \phi}^{*}(\vec x,x^o)\check \phi (\vec x,x^o)-\phi_o^2]-
\nonumber \\
&-&ie^2{\check \pi}_{\phi^{*}}(\vec x,x^o){1\over {\triangle}}[{\check \psi}
^{\dagger}(\vec x,x^o)\check \psi (\vec x,x^o)-i({\check \pi}_{\phi}(\vec x,x^o)
\check \phi (\vec x,x^o)-{\check \phi}_{\phi^{*}}(\vec x,x^o){\check \phi}
^{*}(\vec x,x^o))]\nonumber \\
&&{}\nonumber \\
&&\partial^o{\check \psi}(\vec x,x^o){\buildrel \circ \over =}
\vec \alpha \cdot (\vec \partial -ie{\vec A}_{\perp}(\vec x,x^o))\check \psi
(\vec x,x^o)-im\beta \check \psi (\vec x,x^o)-\nonumber \\
&-&e^2\check \psi(\vec x,x^o){1\over {\triangle}}[{\check \psi}
^{\dagger}(\vec x,x^o)\check \psi (\vec x,x^o)-i({\check \pi}_{\phi}(\vec x,x^o)
\check \phi (\vec x,x^o)-{\check \phi}_{\phi^{*}}(\vec x,x^o){\check \phi}
^{*}(\vec x,x^o))],
\label{16}
\end{eqnarray}

\noindent which imply the following Euler-Lagrange equations

\begin{eqnarray}
\Box {\vec A}_{\perp}(x)&{\buildrel \circ \over =}& 
-e{\check \psi}^{\dagger}(x)\vec \alpha \check \psi (x)-\nonumber \\
&-&ie[{\check \phi}^{*}(x)(\vec \partial -ie{\vec A}_{\perp}(x))\check \phi
(x)-\check \phi (x)(\vec \partial +ie{\vec A}_{\perp}(x)){\check \phi}
^{*}(x)\nonumber \\
{\ddot {\check \phi}}&-&(\vec \partial -ie{\vec A}_{\perp}(x))^2{\check \phi}(x)
{\buildrel \circ \over =}-2\lambda {\check \phi}^{*}(x)[{\check \phi}^{*}(x)
\check \phi (x)-\phi_o^2]-\nonumber \\
&-&ie\lbrace {\dot {\check \phi}}(x)+ie^2\check \phi (x){1\over {\triangle +2e^2
{\check \phi}^{*}(x)\check \phi (x)}}[{\check \psi}^{\dagger}(x)\check \psi
(x)+i({\check \phi}^{*}(x){\dot {\check \phi}}(x)-{\dot {\check \phi}}
^{*}(x)\check \phi (x))]\rbrace -\nonumber \\
&-&ie\partial^o\lbrace \check \phi (x){1\over {\triangle +2e^2
{\check \phi}^{*}(x)\check \phi (x)}}[{\check \psi}^{\dagger}(x)\check \psi
(x)+i({\check \phi}^{*}(x){\dot {\check \phi}}(x)-{\dot {\check \phi}}
^{*}(x)\check \phi (x))]\rbrace \nonumber \\
(\partial^o&-&\vec \alpha \cdot (\vec \partial -ie{\vec A}_{\perp}(x))+im
\beta){\check \psi}(x){\buildrel \circ \over =}\nonumber \\
&{\buildrel \circ \over =}&-e^2\check \psi (x){1\over {\triangle +2e^2
{\check \phi}^{*}(x)\check \phi (x)}}[{\check \psi}^{\dagger}(x)\check \psi
(x)+i({\check \phi}^{*}(x){\dot {\check \phi}}(x)-{\dot {\check \phi}}
^{*}(x)\check \phi (x))].
\label{17}
\end{eqnarray}

\noindent These equations can be recovered from the Lagrangian density of 
Eq.(\ref{15}) by using the identity $f{{\partial}\over {\partial V}}
{1\over {\triangle +V}}f=-f {1\over {(\triangle +V)^2}}f=-{1\over 
{\triangle + V}}f\, {1\over {\triangle + V}}f$ (modulo a surface term),
where $f={\check \psi}^{\dagger}\check \psi =i({\check \phi}^{*}{\dot {\check
\phi}}-{\dot {\check \phi}}^{*}\check \phi )$ and $V=2e^2{\check \phi}^{*}
\check \phi$.
The Higgs field $\phi (x)$ must be such that the operator $\triangle +2e^2
{\check \phi}^{*}(x)\check \phi (x)$ has no zero modes.

Eqs.(\ref{13}) and (\ref{15}) also give the reduction to Dirac's observables of
a charged complex Klein-Gordon field interacting with the electromagnetic field.

\section
{The Higgs phase.}

There are two methods to get this phase starting from the following 
parametrization of the Higgs fields [the value $\phi =0$ is not covered by
these radial coordinates; for the sake of simplicity we take a positive
value $\phi_o > 0$ for the arbitrary symmetry breaking reference point in the 
set of minima of the potential: this set is spanned by varying an angular 
variable $\theta$, so that $\theta$ is the would-be Goldstone boson; the
symmetry group U(1) is broken and there is no residual stability group of the
points of minimum]

\begin{eqnarray}
\phi (x)&=&[\phi_o+{1\over \sqrt{2}}H(x)] e^{ie\theta (x)}={1\over \sqrt{2}}
(v+H(x))e^{ie\theta (x)},\quad\quad v=\sqrt{2}\phi_o,
\nonumber \\
&&D^{(A)}_{\mu}\phi (x)=e^{ie\theta (x)}[{1\over \sqrt{2}}
\partial_{\mu}H(x)-ie(\phi_o+{1\over \sqrt{2}}H(x))
(A_{\mu}(x)-\partial_{\mu}\theta (x))],
\label{18}
\end{eqnarray}

\noindent so that the Lagrangian density becomes

\begin{eqnarray}
{\cal L}(x)&=&-{1\over 4}F_{\mu\nu}(x)F^{\mu\nu}(x)+e^2(\phi_o+
{1\over \sqrt{2}}H(x))^2(A_{\mu}(x)-
\partial_{\mu}\theta (x))(A^{\mu}(x)-\partial^{\mu}\theta (x))+\nonumber \\
&+&{1\over 2}\partial_{\mu}H(x)\partial^{\mu}H(x)-
{{\lambda}\over 2} H^2(x)({1\over \sqrt{2}}H(x)+2\phi_o)^2+\nonumber \\
&+&{1\over 2}\bar \psi (x)[\gamma^{\mu}(i\partial_{\mu}+eA_{\mu}(x))-
\stackrel{\longleftarrow}{(i\partial_{\mu}-eA_{\mu}(x))}\gamma^{\mu}]
\psi (x)-m\bar \psi (x)\psi (x).
\label{19}
\end{eqnarray}

The parametrization of Eq.(\ref{18}) requires a restriction to Higgs
fields which have no zeroes, namely $\phi^{*}(x)\phi (x)\not= 0$ 
[$H(x)\not= -v=-\sqrt{2}\phi_o$], and with
a nonsingular phase $\theta (x)$ because we assumed a trivial U(1)
principal bundle. The analogue of the quantum statement of symmetry
breaking, i.e. that the theory is invariant under a group G but not the
ground state, is replaced by the choice of the parametrization (\ref{18})
with a given $\phi_o$, i.e. by the choice of a family of solutions of
the Euler-Lagrange equations associated with Eq.(\ref{1}) not invariant
under U(1).

i) The canonical momenta coming from Eq.(\ref{19}) are

\begin{eqnarray}
&&\pi^o(x)=0\nonumber \\
&&\vec \pi (x)=\vec E(x)\nonumber \\
&&\pi_H(x)=\partial^oH(x)\nonumber \\
&&\pi_{\theta}(x)=2e^2(\phi_o+{1\over \sqrt{2}}
H(x))^2(\partial_o\theta (x)-A_o(x))\nonumber \\
&&{}\nonumber \\
&&\lbrace H(\vec x,x^o),\pi_H(\vec y,x^o)\rbrace =\lbrace \theta (\vec x,x^o),
\pi_{\theta}(\vec y,x^o)\rbrace =\delta^3(\vec x-\vec y).
\label{20}
\end{eqnarray}

The resulting Dirac Hamiltonian density is

\begin{eqnarray}
{\cal H}_D(x)&=&{1\over 2}[{\vec \pi}^2(x)+{\vec B}^2(x)]+
e^2(\phi_o+{1\over \sqrt{2}}H(x))^2(\vec A(x)-
\vec \partial \theta (x))^2+\nonumber \\
&+&{1\over 2}[\pi_H^2(x)+(\vec \partial H(x))^2]+{{\lambda}\over 2} H^2(x)
({1\over \sqrt{2}}H(x)+2\phi_o)^2+
{{\pi^2_{\theta}(x)}\over {4e^2(\phi_o+{1\over \sqrt{2}}H(x))^2}}+\nonumber \\
&+&\psi^{\dagger}(x)\vec \alpha \cdot (i\vec \partial +
e\vec A(x))\psi (x)+m\bar \psi (x)\psi (x)-\nonumber \\
&-&A_o(x)[-\vec \partial \cdot \vec \pi (x)+e\psi^{\dagger}(x)\psi (x)-
\pi_{\theta}(x)]+\lambda_o(x)\pi^o(x)
\label{21}
\end{eqnarray}

\noindent and there are two first class constraints: $\pi^o(x)\approx 0$ and
the Gauss law

\begin{equation}
\hat \Gamma (x)=-\vec \partial \cdot \vec \pi (x)+e\psi^{\dagger}(x)\psi (x)-
\pi_{\theta}(x)\approx 0,
\label{22}
\end{equation}

\noindent which is now to be 
solved in $\pi_{\theta}(x)$. The pairs of conjugate 
gauge variables are now $A_o(x), \pi^o(x), \theta (x), \hat \Gamma (x)$, while 
the Dirac observables, having zero Poisson bracket with $\hat \Gamma (x)$, are

\begin{eqnarray}
&&{\vec A}^{'}(x)=\vec A(x)-\vec \partial \theta (x),\nonumber \\
&&\vec \pi (x)=\vec E(x),\nonumber \\
&&\hat \psi (x)=e^{-ie\theta (x)}\psi (x),\nonumber \\
&&{\hat \psi}^{\dagger}(x)=e^{ie\theta (x)}\psi^{\dagger}(x),\nonumber \\
&&H(x),\nonumber \\
&&\pi_H(x).
\label{23}
\end{eqnarray}

\noindent and the Coulomb cloud of the electromagnetic phase has been now
replaced by a Higgs (would-be Goldstone boson) cloud, which dresses the fermion
fields and the vector field. In this way the would-be 
Goldstone boson (and the associated infrared singularities at the quantum
level\cite{str}) are ``eaten" by the gauge boson which become massive. This
is connected to the Gauss law\cite{str}, which is not trivial in presence
of spontaneous symmetry breaking with the Higgs mechanism, as we shall see
in the last Section.

After the symplectic decoupling without adding gauge-fixings, we get the
following Hamiltonian density

\begin{eqnarray}
{\cal H}^{(Higgs)}_{phys}(x)&=&{1\over 2}[{\vec \pi}^2(x)+{\vec B}^2(x)]+
{1\over 2}m^2_{em}(1+{{|e|}\over {m_{em}}}H(x))^2{\vec A}^{{'}2}(x)+
\nonumber \\
&+&{1\over 2}[\pi_H^2(x)+(\vec \partial 
H(x))^2]+{1\over 2}
m_H^2\, H^2(x)(1+{{|e|}\over {2m_{em}}}H(x))^2+\nonumber \\
&+&{\hat \psi}^{\dagger}(x)\vec \alpha \cdot
(i\vec \partial +e{\vec A}^{'}(x))\hat \psi (x)+m{\hat {\bar \psi}}(x)\hat 
\psi (x)+{ {(\vec \partial \cdot \vec \pi (x)-e{\hat \psi}^{\dagger}(x)\hat \psi
(x))^2}\over {2m^2_{em}(1+{{|e|}\over {m_{em}}}H(x))^2} },
\label{24}
\end{eqnarray}

\noindent which is local but not analytic in the electric charge e or,
by replacing $\phi_o$ with the electromagnetic mass 
$m_{em}=\sqrt{2}|e|\phi_o=|e|v$, in the sum of the
mass produced by the spontaneous symmetry breaking and the residual Higgs
field, whose mass is $m_H=2\phi_o\sqrt{\lambda}$ [ so that $\phi_o=
m_{em}/\sqrt{2} |e|$ and $\lambda =e^2m_H^2/2m^2_{em}$].
From Eq.(\ref{16}) we
get $\pi_{\theta}=(m_{em}+|e|H)^2(\partial_o\theta +A_o)$.
Let us remark that in those points $x^{\mu}$ where $H(x)=-m_{em}/|e|=-
\sqrt{2}\phi_o$ [which were excluded to exist not to have problems
with the origin of the radial coordinates of Eq.(\ref{18})]
we would recover massless electromagnetism,
so that the numerator of the self-energy term in Eq.(\ref{24}) must vanish,
being the Gauss law of the massless theory. Therefore we should not have a
singularity in these points, but new physical effects as shown in Section IV.

Let us remark that the self-energy appearing in Eq.(\ref{24}) is local and
that, in presence of fermion fields, it contains a 4 fermion interaction,
which has appeared from the nonperturbative solution of the Gauss law and
which is a further obstruction to the renormalizability of the reduced theory
(equivalent to the unitary gauge, but without having added any gauge-fixing),
which already fails in the unitary physical gauge due to the massive vector
boson propagator not fulfilling the power counting rule; as said
in Ref.\cite{or}, this is due to the fact that the field-dependent gauge 
transformation relating $\vec A$ and ${\vec A}^{'}$ in Eq.(\ref{23}) is not
unitarily implementable. It is interesting to note that all the interaction
terms of the residual Higgs field $H(x)$ in Eq.(\ref{24}) show that it
couples to the ratio $|e|/m_{em}$.

Again the lack of manifest
Lorentz covariance can be taken care of by reformulating the theory on
spacelike hypersurfaces, as shown in Section IV.

Since one has

\begin{eqnarray}
\partial^oA^{{'}i}(\vec x,x^o)&=&\lbrace A^{{'}i}(\vec x,x^o),\int d^3y{\cal H}
^{(Higgs)}_{phys}(\vec y,x^o)\rbrace =\nonumber \\
&=&-\pi^i(\vec x,x^o)+\partial^i\, {{\vec \partial \cdot \vec \pi (\vec x,x^o)-
e{\hat \psi}^{\dagger}(\vec x,x^o)\hat \psi (\vec x,x^o)}\over {(m_{em}+
|e|H(x))^2}}\nonumber \\
&&{}\nonumber \\
\Rightarrow &\pi^i&(x)=-\partial^oA^{{'}i}(x)-\nonumber \\
&-&\partial^i\, {1\over {\triangle +(m_{em}+|e|H(x))^2}}\, [\vec \partial \cdot
\partial^o{\vec A}^{'}(x)+e{\hat \psi}^{\dagger}(x)\hat \psi (x)],
\label{25}
\end{eqnarray}

\noindent we get a nonlocal Lagrangian density describing only the Higgs
phase

\begin{eqnarray}
{\cal L}^{(Higgs)}_{phys}(x)&=&{\hat \psi}^{\dagger}(x)[i\partial^o-\vec
\alpha \cdot (i\vec \partial +e{\vec A}^{'}(x))-\beta m]\hat \psi (x)+
\nonumber \\
&+&{1\over 2}[(\partial^o{\vec A}^{'}(x))^2-[\vec \partial \cdot \partial^o{\vec
A}^{'}(x)+e{\hat \psi}^{\dagger}(x)\hat \psi (x)]\nonumber \\
&&{1\over {\triangle +
m^2_{em}(1+{|e|\over {m_{em}}}H(x))^2}}[\vec \partial \cdot \partial^o{\vec
A}^{'}(x)+e{\hat \psi}^{\dagger}(x)\hat \psi (x)]]-{1\over 2}{\vec B}^2(x)
-\nonumber \\
&-&{1\over 2}m^2_{em}(1+{{|e|}\over {m_{em}}}H(x))^2{\vec A}^{{'}2}(x)
+\nonumber \\
&+&{1\over 2}\partial_{\mu}H(x)\partial^{\mu}H(x)-{1\over 2}
m^2_H H^2(x)(1+{{|e|}\over {2m_{em}}}
H(x))^2.
\label{26}
\end{eqnarray}

We see that the potential problems of the Hamiltonian formulation at the 
points where $H(x)=-m_{em}/|e|=-\sqrt{2}\phi_o$ are now replaced bythe
requirement that the operator $\triangle +m_{em}^2(1+{{|e|}\over {m_{em}}}
H(x))^2$ must not have zero modes.

ii) Since the Dirac observables ${\vec A}^{'}=\vec A-\vec \partial \theta$ are
obtained from $\vec A$ with a $\theta$-field-dependent gauge transformation,
which is the space part of the gauge transformation $A_{\mu}\mapsto A^{'}
_{\mu}=A_{\mu}-\partial_{\mu}\theta$ , $\phi (x)\mapsto \phi_o$, $\psi (x)
\mapsto \hat \psi (x)=e^{-ie\theta (x)}\psi (x)$, used to go to the ``unitary 
gauge"\cite{hoo,cl}, we can do this gauge transformation in the gauge invariant
Lagrangian density of Eq.(\ref{19}) to get

\begin{eqnarray}
{\cal L}(x)&=&{\cal L}^{'}(x)=\nonumber \\
&=&-{1\over 4}F_{\mu\nu}(x)F^{\mu\nu}(x)+{1\over 2}
m^2_{em}(1+{{|e|}\over {m_{em}}}H(x))^2A^{{'}2}(x)+\nonumber \\
&+&{1\over 2}\partial_{\mu}H(x)\partial^{\mu}H(x)-{1\over 2}
m_H^2 H^2(x)(1+{{|e|}\over {2m_{em}}}
H(x))^2+\nonumber \\
&+&{\hat \psi}^{\dagger}(x)\gamma^{\mu}(i\partial_{\mu} +eA^{'}_{\mu}(x))\hat 
\psi (x)-m{\hat {\bar \psi}}(x)\hat \psi (x).
\label{27}
\end{eqnarray}

Now ${\cal L}^{'}(x)$ depends on $A^{'}_{\mu}, H, \hat \psi , {\hat \psi}
^{\dagger}$, but not on $\theta$. The new momenta are

\begin{eqnarray}
&&\pi^o(x)=0,\nonumber \\
&&\vec \pi (x)=\vec E(x),\nonumber \\
&&\pi_H(x)=\partial^oH(x),
\label{28}
\end{eqnarray}

\noindent and the new Dirac Hamiltonian density is

\begin{eqnarray}
{\cal H}^{'}_D(x)&=&{1\over 2}[{\vec \pi}^2(x)+{\vec B}^2(x)]-
{1\over 2}m^2_{em}(1+{{|e|}\over {m_{em}}}H(x))^2
[A^{{'}2}_o(x)-{\vec A}^{{'}2}(x)]+\nonumber \\
&+&{\hat \psi}^{\dagger}(x)\vec \alpha \cdot (i\vec \partial +e{\vec A}^{'}
(x))\hat \psi (x)+m{\hat {\bar \psi}}(x)\hat \psi (x)-\nonumber \\
&-&A^{'}_o(x)[-\vec \partial \cdot \vec \pi (x)+e{\hat \psi}^{\dagger}(x)
\hat \psi (x)]+{1\over 2}[\pi_H^2(x)+(\vec \partial H(x))^2]+\nonumber \\
&+&{1\over 2}
m_H^2 H^2(x)(1+{{|e|}\over {2m_{em}}}H(x))^2+\lambda_o(x)\pi^o(x).
\label{29}
\end{eqnarray}

Now the time constancy of the primary constraint $\pi^o(x)\approx 0$
generates the $A^{'}_o$-dependent secondary constraint

\begin{equation}
\zeta (x)=(m_{em}+|e|H(x))^2A^{'}_o(x)-\vec \partial \cdot \vec \pi (x)+
e{\hat \psi}^{\dagger}(x)\hat \psi (x)\approx 0.
\label{30}
\end{equation}

The time constancy of $\zeta (x)\approx 0$ determines the Dirac multiplier
$\lambda_o(x)$, so that now $\pi^o(x)\approx 0, \zeta (x)\approx 0$ are a
pair of second class constraints eliminating $A^{'}_o(x)$ and $\pi^o(x)$
by going to Dirac brackets. The substitution of the value of $A^{'}_o(x)$
given by $\zeta (x)\equiv 0$ into Eq.(\ref{29}) reproduces Eq.(\ref{24}).

Let us remark that this mechanism of second class constraints is the same
which acts in the search of Dirac's observables of the standard massive vector 
field described by the Lagrangian density 

\begin{equation}
{\cal L}(x)=-{1\over 4}F_{\mu\nu}(x)F^{\mu\nu}(x)
+{1\over 2}M^2A_{\mu}(x)A^{\mu}(x),
\label{31}
\end{equation} 

\noindent which is not gauge invariant under
U(1) local gauge transformations. Its Euler-Lagrange equations $\partial_{\nu}
F^{\nu\mu}(x)+M^2A^{\mu}(x)\, {\buildrel \circ \over =}\, 0$ imply
$(\Box +M^2)A_{\mu}(x)\, {\buildrel \circ \over =}\, 0$ and $\partial^{\mu}
A_{\mu}(x){\buildrel \circ \over =}\, 0$. The canonical momenta are $\pi^o(x)=0$
and $\vec \pi (x)=\vec E(x)$. Associated with the primary constraint
$\pi^o(x)\approx 0$ there is the following gauge transformation $\delta A_o(x)=
\Lambda (x)$ [$\Lambda (x)$ arbitrary function], $\delta \vec A(x)=0$, under
which the Lagrangian density is quasi-invariant, $\delta {\cal L}(x)=(\partial_k
F^{ko}(x)+M^2A^o(x))\delta A_o(x)\, {\buildrel \circ \over =}\, 0$, as it must
be with second-class primary constraints\cite{lu,lus}.
The canonical Dirac Hamiltonian density
is ${\cal H}_D(x)={1\over 2}[{\vec \pi}^2(x)+{\vec B}^2(x)]-{{M^2}\over 2}
[A_o^2(x)-{\vec A}^2(x)]+A_o(x)\vec \partial \cdot \vec \pi (x)+
\lambda_o(x)\pi^o(x)$. The time constancy of $\pi^0(x)\approx 0$
generates the secondary constraint $\zeta^{'}(x)=M^2A_o(x)-\vec \partial \cdot
\vec \pi (x)\approx 0$ and its time constancy determines the Dirac multiplier
$\lambda_o(x)\, {\buildrel \circ \over =}\, \vec \partial \cdot \vec A(x)$
[i.e. $\partial^{\mu}A_{\mu}(x)\, {\buildrel \circ \over =}\, 0$, because
$\lambda_o(x)\, {\buildrel \circ \over =}\, \partial^oA_o(x)$]. 
The Dirac observables of the model are $\vec A(x), \vec \pi
(x)$, and the final physical Hamiltonian and Lagrangian densities are

\begin{eqnarray}
{\cal H}^{(mass)}_{phys}(x)&=&{1\over 2}[{\vec \pi}^2(x)+{\vec B}^2(x)]+
{1\over 2}M^2{\vec A}^2(x)+{ {(\vec \partial \cdot \vec \pi (x))^2}\over
{2M^2}}\nonumber \\
&&{}\nonumber \\
{\cal L}^{(mass)}_{phys}(x)&=&{1\over 2}[(\partial^o\vec A(x))^2-
\vec \partial \cdot \partial^o\vec A(x) {1\over {\triangle +M^2}}
\vec \partial \cdot \partial^o\vec A(x)]-\nonumber \\
&-&{1\over 2}{\vec B}^2(x)-{1\over 2}M^2 {\vec A}^2(x).
\label{32}
\end{eqnarray}

From the Hamilton equations ${\dot A}^i(x)\, {\buildrel \circ \over =}\, -
(\delta^{ij}-{{\partial^i\partial^j}\over {M^2}})\pi^j(x)$, ${\dot \pi}^i(x)
{\buildrel \circ \over =}\, (\triangle +M^2)(\delta^{ij}+{{\partial^i
\partial^j}\over {\triangle +M^2}})A^j(x)$,
we get $\pi^i=-(\delta^{ij}+{{\partial^i\partial^j}
\over {\triangle +M^2}})\partial^oA^j$ and the Euler-Lagrange equations

\begin{eqnarray}
&&(\Box +M^2)(\delta^{ij}+{{\partial^i\partial^j}
\over {\triangle +M^2}}) A^j(x){\buildrel \circ \over =} 0\nonumber \\
\Rightarrow && {1\over {\triangle +M^2}}(\Box +M^2)\vec \partial \cdot \vec A
(x)\, {\buildrel \circ \over =}\, 0,\quad\quad (\Box +M^2)\vec A(x){\buildrel
\circ \over =}\, 0.
\label{m1}
\end{eqnarray}

As noted in Ref.\cite{kks}, one can consistently eliminate the residual Higgs
field $H(x)$ at the Hamiltonian level, 
even if in the Abelian theory it is physically relevant being
connected to amplitude effects coming from the condensate (Cooper pairs)
simulated by the scalar fields. The elimination can be done by adding to the
Hamiltonian density of the Higgs phase, Eq.(\ref{24}), or to the Lagrangian
density (\ref{26}), a term $\mu (x)H(x)$ with $\mu (x)$ a Lagrange
multiplier. This would imply a new holonomic constraint $H(x)\approx 0$
[in some sense it would correspond to $m_H\rightarrow \infty$; 
one could also require $H(x)-h\approx 0$ with $h=const.$]
whose time constancy at the Hamiltonian level would generate the secondary
constraint $\pi_H(x)\approx 0$. The time constancy of this secondary constraint
would determine the multiplier $\mu (x)$, so that the two constraints turn
out to be second class. By going to Dirac brackets, we would obtain a theory
without residual Higgs fields described by the Hamiltonian density of Eq.
(\ref{24}) evaluated at $H(x)=\pi_H(x)\equiv 0$

\begin{eqnarray}
{\tilde {\cal H}}^{(Higgs)}_{phys}(x)
&=&{1\over 2}[{\vec \pi}^2(x)+{\vec B}^2(x)]+
{1\over 2}m^2_{em}{\vec A}^{{'}2}(x)+
{\hat \psi}^{\dagger}(x)\vec \alpha \cdot
(i\vec \partial +e{\vec A}^{'}(x))\hat \psi (x)+\nonumber \\
&+&m{\hat {\bar \psi}}(x)\hat \psi (x)+
{ {(\vec \partial \cdot \vec \pi (x)-e{\hat \psi}^{\dagger}(x)\hat \psi
(x))^2}\over {2m_{em}^2} }.
\label{33}
\end{eqnarray}

\noindent to be compared with Eq.({\ref{32}). The elimination of $H(x)$
reproduces the massive vector theory and can also be thought
as a limiting classical result of the so-called ``triviality problem"
[triviality of the $\lambda \phi^4$ theory \cite{tri}], which however would 
imply a quantization (but how?) of the
Higgs phase alone without the residual Higgs field, so that also its
quantum fluctuations would be absent (instead they are the main left
quantum effect in the limit $m_H\rightarrow \infty$, which is known to
produce\cite{ab}, in the non-Abelian case, a gauge theory coupled to a
nonlinear $\sigma$-model, equivalent\cite{bs} to a massive Yang-Mills
theory).

The physical Hamiltonian of Eq.(\ref{26}) implies the Hamilton equations

\begin{eqnarray}
\partial^o{\vec A}^{'}
(\vec x,x^o)&{\buildrel \circ \over =}& -\vec \pi (\vec x,x^o)
+\vec \partial \,\,
{ {\vec \partial \cdot \vec \pi (\vec x,x^o)-e{\hat \psi}^{\dagger}(\vec x,x^o)
\hat \psi (\vec x,x^o)}\over {(m_{em}+|e|H(\vec x,x^o))^2}}\nonumber \\
\partial^o\vec \pi (\vec x,x^o)&{\buildrel \circ \over =}& (m_{em}+|e|H(\vec x,
x^o))^2{\vec A}^{'}(\vec x,x^o)+e{\hat \psi}^{\dagger}(\vec x,x^o)\vec \alpha
\hat \psi (\vec x,x^o)+\nonumber \\
&+&\triangle {\vec A}^{'}(\vec x,x^o)+\vec\partial
(\vec \partial \cdot {\vec A}^{'}(\vec x,x^o))\nonumber \\
\partial^oH(\vec x,x^o)&{\buildrel \circ \over =}& \pi_H(\vec x,x^o)\nonumber \\
\partial^o\pi_H(\vec x,x^o)&{\buildrel \circ \over =}& -|e|(m_{em}+|e|H
(\vec x,x^o)){\vec A}^{{'}2}(\vec x,x^o)-\triangle H(\vec x,x^o)-\nonumber \\
&-&m_H^2
H(\vec x,x^o)(1+{{|e|}\over {2m_{em}}}H(\vec x,x^o))^2-
{{|e|m_H^2}\over {2m_{em}}}H^2(\vec x,x^o)(1+{{|e|}\over {2m_{em}}}
H(\vec x,x^o))+\nonumber \\
&+&|e|{{(\vec \partial \cdot \vec \pi (\vec x,x^o)-e{\hat \psi}^{\dagger}
(\vec x,x^o)\hat \psi (\vec x,x^o))^2}\over {(m_{em}+|e|H(\vec x,x^o))^3}}
\nonumber \\
\partial^o{\hat \psi}(\vec x,x^o)&{\buildrel \circ \over =}&\vec \alpha \cdot
(\vec \partial -ie{\vec A}^{'}(\vec x,x^o))\hat \psi (\vec x,x^o)-im\gamma^o\hat
\psi (\vec x,x^o)+\nonumber \\
&+&ie\hat \psi (\vec x,x^o)\,\, {{\vec \partial \cdot \vec \pi
(\vec x,x^o)-e{\hat \psi}^{\dagger}(\vec x,x^o)\hat \psi (\vec x,x^o)}\over
{(m_{em}+|e|H(\vec x,x^o))^2} }.
\label{34}
\end{eqnarray}

From them we get ${{\vec \partial \cdot \vec \pi -e{\hat \psi}^{\dagger}\hat
\psi}\over {(m_{em}+|e|H)^2}}=-{1\over {\triangle +(m_{em}+|e|H)^2}}
(\partial^o\vec \partial \cdot {\vec A}^{'}+e{\hat \psi}^{\dagger}\hat \psi)$,
$\vec \pi =-\partial^o{\vec A}^{'}-\vec \partial \, {1\over {\triangle +
(m_{em}+|e|H)^2}}
(\partial^o\vec \partial \cdot {\vec A}^{'}+e{\hat \psi}^{\dagger}\hat \psi)$
and the following Euler-Lagrange equations

\begin{eqnarray}
\lbrace \Box &+&(m_{em}+|e|H)^2\rbrace {\vec A}^{'}(x)+[\vec \partial +
\partial^o{1\over {\triangle +(m_{em}+|e|H)^2}}\partial^o]\vec \partial \cdot
{\vec A}^{'}(x)+\nonumber \\
&+&{(m_{em}+|e| H(x))}^2{\vec A}^{'}(x){\buildrel \circ \over =}
-e{\hat \psi}^{\dagger}(x)\vec \alpha \hat \psi (x)-
\partial^o\vec \partial {1\over {\triangle +(m_{em}+|e|H)^2}}(e{\hat 
\psi}^{\dagger}(x)\hat \psi (x))\nonumber \\
\Box H(x) &{\buildrel \circ \over =}& -|e| (m_{em}+|e| H(x)){\vec A}^{{'}2}(x)-
m^2_HH(x){(1+{ {|e|}\over {2m_{em}} }H(x))}^2-\nonumber \\
&-&{ {|e| m_H^2}\over {2m_{em}} }H^2(x)(1+{ {|e|}\over {2m_{em}} }H(x))+
\nonumber \\
&+&|e| (m_{em}+|e|H(x))[{1\over {\triangle +(m_{em}+|e|H)^2}}(\partial^o\vec 
\partial \cdot {\vec A}^{'}(x)+e{\hat 
\psi}^{\dagger}(x)\hat \psi (x))]{}^2
\nonumber \\
( i\partial^o &-&
\vec \alpha \cdot (i\vec \partial +e{\vec A}^{'}(x))-m\gamma^o)
\hat \psi (x){\buildrel \circ \over =}\nonumber \\
&{\buildrel \circ \over =}& e\hat \psi (x)
{1\over {\triangle +(m_{em}+|e|H)^2}}(\partial^o\vec \partial \cdot 
{\vec A}^{'}(x)+e{\hat 
\psi}^{\dagger}(x)\hat \psi (x)).
\label{35}
\end{eqnarray}

\noindent which can be recovered from the Lagrangian density of Eq.(\ref{23})
by using the same identity given at the end of Section 2 with $f=\partial^o
\vec \partial \cdot {\vec A}^{'}+e{\hat \psi}^{\dagger}\hat \psi$ and
$V=(m_{em}+|e|H)^2$.

We do not know how to solve the coupled Eqs.(\ref{35}). Let us make two comments
on the case without fermions, so that the third line of Eqs.(\ref{35}) is
missing. By turning off the coupling constant $e$ the first of Eqs.(\ref{35})
becomes consistent with the massive vector theory, while the 
second one becomes a Klein-Gordon equation for $H(x)$. By putting $H(x)=0$ in 
Eqs.(\ref{35}) (limit $m_H\rightarrow \infty$), the first line still reproduces
the massive vector theory, but the second line becomes the
restriction $|e|m_{em}[({{\partial^o}\over {\triangle +m_{em}^2}}\vec
\partial \cdot {\vec A}^{'})^2-{\vec A}^{{'}2}]{\buildrel
\circ \over =} 0$ on the space of solutions of Eq.(\ref{m1}) [in this
approximation, with ${\vec A}^{'}={\vec A}^{'}_{\perp}-{{\vec \partial}\over
{\triangle}} \vec \partial \cdot {\vec A}^{'}$, the first of Eqs.(\ref{35})
becomes $(\Box +m^2_{em}){\vec A}^{'}_{\perp}{\buildrel \circ \over =} 0$,
$(\partial_o^2{{m_{em}^2}\over {\triangle +m^2_{em}}}+ m_{em}^2)\vec
\partial \cdot {\vec A}^{'} {\buildrel \circ \over =} 0$]. Therefore a
weak nearly constant Higgs field (strongly interacting symmetry breaking
sector for $m_h\rightarrow \infty$) influences the longitudinal polarization
of the massive vector field; this is the main difference  from the massive 
vector theory introduced by the Higgs mechanism.

Let us remark that, while in the electromagnetic phase the self-energy term
in the physical Hamiltonian (\ref{12}) is nonlocal, the self-energy term
in Eq.(\ref{24}) of the Higgs phase is local implying a local four-fermion
coupling. This feature is common to the physical Hamiltonian (\ref{32}) of
the standard massive vector field and to Eq.(\ref{33}). There is a
not-manifestly Lorentz covariant modification of Eq.(\ref{31}) involving
only the nonphysical variable $A_o(x)$, namely

\begin{equation}
{\cal L}^{'}(x)=-{1\over 4}F_{\mu\nu}(x)F^{\mu\nu}(x)+{1\over 2}[M^2A_o^2(x)+
\vec \partial A_o(x)\cdot \vec \partial A_o(x)]-{1\over 2}M^2{\vec A}^2(x),
\label{m2}
\end{equation}

\noindent which solves this problem and which can be made Lorentz covariant
by its reformulation on spacelike hypersurfaces as shown in Section V.
The new Euler-Lagrange equations and the Dirac Hamiltonian density are 
respectively [the canonical momenta are the same of the massive vector theory]

\begin{eqnarray}
&&\partial_{\nu}F^{\nu\mu}(x)+M^2A^{\mu}(x)+\eta^{\mu o}\triangle A^o(x)
{\buildrel \circ \over =}\, 0\nonumber \\
&&\Rightarrow \quad (\Box +M^2)A^{\mu}(x)-\partial^{\mu}\partial_{\nu}A^{\nu}
(x)+\eta^{\mu o}\triangle A^o(x)\, {\buildrel \circ \over =}\, 0\nonumber \\
&&{}\nonumber \\
&&{\cal H}^{'}(x)={1\over 2}[{\vec \pi}^2(x)+{\vec B}^2(x)]-{1\over 2}A_o(x)
(\triangle +M^2)A_o(x)+{1\over 2}M^2{\vec A}^2(x)+\nonumber \\
&&+A_o(x)\vec \partial \cdot \vec \pi (x)+\lambda_o(x)\pi^o(x).
\label{m3}
\end{eqnarray}

The time constancy of the primary constraint $\pi^o(x)\approx 0$ produces the
secondary one $\tilde \zeta (x)=(\triangle +M^2)A_o(x)-\vec \partial \cdot
\vec \pi (x)\approx 0$, whose time constancy determines $\lambda_o(x)\approx
{{M^2}\over {\triangle +M^2}}\vec \partial \cdot \vec A(x)$ consistently
with the Euler-Lagrange equations and with $\lambda_o(x)\, {\buildrel \circ
\over =}\, {\dot A}_o(x)$. By eliminating $A_o(x), \pi^o(x)$ with the pair
of second class constraints, we arrive at the physical Hamiltonian density

\begin{equation}
{\cal H}^{'}_{phys}(x)={1\over 2}[{\vec \pi}^2(x)+{\vec B}^2(x)]+{1\over 2}M^2
{\vec A}^2(x)+{1\over 2}\vec \partial \cdot \vec \pi (x)\, {1\over {\triangle +
M^2}}\, \vec \partial \cdot \vec \pi (x).
\label{m4}
\end{equation}

A gauge transformation $\delta A_o(x)$, generated by $\pi^o(x)\approx 0$, 
would produce a weak quasi-invariance\cite{lus3} $\delta {\cal L}(x)=-
[\partial_{\nu}F^{\nu o}(x)+(\triangle +M^2)A^o(x)]\delta A_o(x)=-[(2
\triangle +M^2)A^o(x)+\partial^o\vec \partial \cdot \vec A(x)]\delta A_o(x)
{\buildrel \circ \over =} 0$, i.e. $\delta {\cal L}(x)$ vanishes by using the
acceleration independent Euler-Lagrange equation corresponding to the Gauss
law.

The Hamilton equations ${\dot A}^i(x)\, {\buildrel \circ \over =}\, -\pi^i(x)+
{{\partial^i}\over {\triangle +M^2}}\, \vec \partial \cdot \vec \pi (x)$,
${\dot \pi}^i(x)\, {\buildrel \circ \over =}\, (\triangle +M^2)A^i(x)+
\partial^i\vec \partial \cdot \vec A(x)$, imply $\pi^i(x)=-(\delta^{ij}+
{{\partial^i\partial^j}\over {2\triangle +M^2}}){\dot A}^j(x)$ and

\begin{eqnarray}
&&(\Box +M^2)A^i(x)+\partial^i[{{\partial_o^2}\over {2\triangle +M^2}}+1]
\vec \partial \cdot \vec A(x)\, {\buildrel \circ \over =}\, 0\nonumber \\
&&\Rightarrow \quad {1\over {2\triangle +M^2}}[(\triangle +M^2)(\Box +M^2)
-\triangle^2] \vec \partial \cdot \vec A(x)\, {\buildrel \circ \over =}\, 0,
\quad\quad (\Box +M^2)A^i_{\perp}(x)\, {\buildrel \circ \over =}\, 0,
\label{m5}
\end{eqnarray}

\noindent where $A^i=A^i_{\perp}-{{\partial^i}\over {\triangle}}\vec \partial \cdot
\vec A$.

In this way one gets a nonlocal self-energy (avoiding a local four-fermion
interaction when fermions are present) with the correct massive Green function
$e^{-M|\, \vec x-\vec y\, |}/4\pi |\, \vec x-\vec y\,|$.
The transverse field still obeys the wave equation, while the longitudinal
field has a modified wave equation [it can also be written in the form
$[\partial^2_o{{\triangle +M^2}\over {2\triangle +M^2}}+M^2]\vec \partial \cdot
\vec A(x)\, {\buildrel \circ \over =}\, 0$].

Therefore, both the Higgs field, Eq.(\ref{35}) without fermions, and this
modification of the standard massive theory produce complicated equations of
motion for the longitudinal part of the vector field.

Finally, to introduce a similar effect in the Lagrangian density (\ref{1}),
we should add a term ${1\over 2} \vec \partial A^{'}_o(x) \cdot \vec \partial 
A^{'}_o(x)$ to the Lagrangian density (\ref{27}) in the unitary gauge,
because in this way the secondary constraint (\ref{30}) would become $\zeta (x)
=[\triangle +(m_{em}+|e| H(x))^2] A^{'}_o(x) - \vec \partial \cdot \vec \pi (x)
+e{\hat \psi}^{\dagger}(x)\hat \psi (x) \approx 0$ and the last term in the
physical Hamiltonian (\ref{24}) would be replaced by ${1\over 2} [\vec \partial
\cdot \vec \pi (x)-e{\hat \psi}^{\dagger}(x)\hat \psi (x)] {1\over {\triangle +
m^2_{em}(1+{{|e|}\over {m_{em}}} H(x))^2}} [\vec \partial
\cdot \vec \pi (x)-e{\hat \psi}^{\dagger}(x)\hat \psi (x)]$ avoiding the local 
4-fermion interaction. Therefore, the Lagrangian density (\ref{1}) should be
replaced by ${\cal L}(x)+{1\over 2} \vec \partial A_o(x) \cdot \vec \partial
A_o(x)$, and we get the following modification of Eqs.(\ref{34}), (\ref{35})

\begin{eqnarray}
\partial^o{\vec A}^{'}(\vec x,x^o)&{\buildrel \circ \over =}& -\vec \pi 
(\vec x,x^o)+\vec \partial \,\,
{ 1\over {\triangle +(m_{em}+|e|H(\vec x,x^o))^2}}\, [\vec \partial \cdot \vec 
\pi (\vec x,x^o)-e{\hat \psi}^{\dagger}(\vec x,x^o)\hat \psi (\vec x,x^o)\, ]
\nonumber \\
\partial^o\vec \pi (\vec x,x^o)&{\buildrel \circ \over =}& (m_{em}+|e|H(\vec x,
x^o))^2{\vec A}^{'}(\vec x,x^o)+e{\hat \psi}^{\dagger}(\vec x,x^o)\vec \alpha
\hat \psi (\vec x,x^o)+\nonumber \\
&+&\triangle {\vec A}^{'}(\vec x,x^o)+\vec\partial
(\vec \partial \cdot {\vec A}^{'}(\vec x,x^o))\nonumber \\
\partial^oH(\vec x,x^o)&{\buildrel \circ \over =}& \pi_H(\vec x,x^o)\nonumber \\
\partial^o\pi_H(\vec x,x^o)&{\buildrel \circ \over =}& -|e|(m_{em}+|e|H
(\vec x,x^o)){\vec A}^{{'}2}(\vec x,x^o)-\triangle H(\vec x,x^o)-\nonumber \\
&-&m_H^2
H(\vec x,x^o)(1+{{|e|}\over {2m_{em}}}H(\vec x,x^o))^2-
{{|e|m_H^2}\over {2m_{em}}}H^2(\vec x,x^o)(1+{{|e|}\over {2m_{em}}}
H(\vec x,x^o))+\nonumber \\
&+&|e|(m_{em}+|e|H(\vec x,x^o))\, [\vec \partial \cdot \vec \pi (\vec x,x^o)-
e{\hat \psi}^{\dagger}(\vec x,x^o)\hat \psi (\vec x,x^o)]\nonumber \\
&&{1\over {(\triangle +(m_{em}+|e|H(\vec x,x^o))^2)^2}}
[\vec \partial \cdot \vec \pi (\vec x,x^o)-e{\hat \psi}^{\dagger}
(\vec x,x^o)\hat \psi (\vec x,x^o)]\nonumber \\
\partial^o{\hat \psi}(\vec x,x^o)&{\buildrel \circ \over =}&\vec \alpha \cdot
(\vec \partial -ie{\vec A}^{'}(\vec x,x^o))\hat \psi (\vec x,x^o)-im\gamma^o\hat
\psi (\vec x,x^o)+\nonumber \\
&+&ie(m_{em}+|e|H(\vec x,x^o))\hat \psi (\vec x,x^o)\nonumber \\
&& {1\over{\triangle +(m_{em}+|e|H(\vec x,x^o))^2} }\,
[\vec \partial \cdot \vec \pi
(\vec x,x^o)-e{\hat \psi}^{\dagger}(\vec x,x^o)\hat \psi (\vec x,x^o)]
\nonumber \\
&&{}\nonumber \\
\lbrace \Box &+&(m_{em}+|e|H)^2\rbrace {\vec A}^{'}(x)+\vec \partial \, 
\vec \partial \cdot {\vec A}^{'}(x){\buildrel \circ \over =}\nonumber \\
&{\buildrel \circ \over =}&-e{\hat \psi}^{\dagger}(x)\vec \alpha \hat \psi (x)-
\partial^o\vec \partial {{\partial^o \, \vec \partial \cdot {\vec A}^{'}(x)+
e{\hat \psi}^{\dagger}(x)\hat \psi (x)}\over {(m_{em}+|e|H)^2}}    \nonumber \\
\Box H(x) &{\buildrel \circ \over =}&-|e| (m_{em}+|e| H(x)){\vec A}^{{'}2}(x)-
m^2_HH(x){(1+{ {|e|}\over {2m_{em}} }H(x))}^2-\nonumber \\
&-&{ {|e| m_H^2}\over {2m_{em}} }H^2(x)(1+{ {|e|}\over {2m_{em}} }H(x))+
\nonumber \\
&+&|e| (m_{em}+|e|H(x))[{1\over {\triangle +(m_{em}+|e|H)^2}}
{{\partial^o\vec \partial \cdot {\vec A}^{'}(x)+e{\hat \psi}^{\dagger}(x)\hat 
\psi (x))}\over {(m_{em}+|e|H(x))^2}}]\nonumber \\
&& (\triangle +(m_{em}+|e|H(x))^2)^2
[{1\over {\triangle +(m_{em}+|e|H)^2}}
{{\partial^o\vec \partial \cdot {\vec A}^{'}(x)+e{\hat \psi}^{\dagger}(x)\hat 
\psi (x))}\over {(m_{em}+|e|H(x))^2}}]\nonumber \\
( i\partial^o &-&
\vec \alpha \cdot (i\vec \partial +e{\vec A}^{'}(x))-m\gamma^o)
\hat \psi (x){\buildrel \circ \over =}\nonumber \\
&{\buildrel \circ \over =}& e\hat \psi (x) 
{{\partial^o\vec \partial \cdot {\vec A}^{'}(x)+e{\hat \psi}^{\dagger}(x)\hat 
\psi (x))}\over {m_{em}+|e|H(x)}}.
\label{m6}
\end{eqnarray}

In the weak nearly constant Higgs field approximation, the analogue of the first
of Eqs.(\ref{35}), with ${\vec A}^{'}={\vec A}^{'}_{\perp}-{{\vec \partial}
\over {\triangle}} \vec \partial \cdot {\vec A}^{'}$, becomes $(\Box +m_{em}^2)
{\vec A}^{'}_{\perp}  {\buildrel \circ \over =} 0$, $[\partial^2_o {{\triangle
+m_{em}^2}\over {2\triangle +m^2_{em}}} +m_{em}^2] \vec \partial \cdot {\vec A}
^{'} {\buildrel \circ \over =} 0$, while the second of Eqs.(\ref{35}) gives
the restriction $|e| m_{em} [ {{\partial^o}\over {\triangle +m_{em}^2}}\vec
\partial \cdot {\vec A}^{'}\, (1+{{\triangle}\over {m_{em}^2}})^2\, {{\partial
^o}\over {\triangle +m_{em}^2}}\vec \partial \cdot {\vec A}^{'} -{\vec A}^{{'}
 2}] {\buildrel \circ \over =} 0$.

\section
{Nielsen-Olesen vortices}

In this paper we have considered only trivial U(1) principal bundles over
Minkowski spacetime (or better over its 
fixed $x^o$ slices $R^3$), avoiding monopole
configurations\cite{go}. As shown for instance in Ref.\cite{bala}, in presence 
of monopoles one has a nontrivial U(1) principal bundle over $M^3=R^3-
\lbrace set\, of\, points\, where\ monopoles\, are\, located \rbrace$
[so that $\pi_o(M^3)=\pi_1(M^3)=0$ but $\pi_2(M^3)\not= 0$, with $\pi_k(M^3)$
being the k-th homotopy group of $M^3$]. Therefore the gauge potentials 
$A_{\mu}$ (cross sections of the U(1) principal bundle) and the Lagrangian 
density of Eq.(\ref{1}) cannot be globally defined, since there are no
global cross sections; instead there is a well defined Hamiltonian
formalism. In our formalism potential problems in the Higgs phase arise
in those points where $H(x)=-m_{em}/|e|=-\sqrt{2}\phi_o$, in which the theory
is nonanalytic and could have, a priori, essential singularities. See
Ref.\cite{and} for a possible generation of mass in the Abelian Higgs
model based not on the Higgs mechanism but on the requirement of 
integrability of the equations of motion and of absence of essential
singularities. In any case, the existence of zeroes $H(x)=-m_{em}/|e|$
is compatible with the existence, in the framework of monopoles, of
static finite-energy solutions with a non-trivial behaviour at space
infinity\cite{go} in the case of two dimensions (Nielsen-Olesen vortices
\cite{no}), whose approximate existence in 3+1 dimensions is welcome
for the theory of superconductivity.

In a type I superconductor the ratio $\kappa =\tilde \lambda /\xi=
m_H/\sqrt{2}m_{em}$ of the magnetic field penetration depth $\tilde 
\lambda =1/\sqrt{2}m_{em}$ and of the coherence length $\xi =1/m_H$ 
[it gives a scale for the variations of the order parameter $\phi (x)$]
satisfies $\sqrt{2} \kappa \leq 1$, which corresponds to $m_H\leq m_{em}$
in the simulation of the Ginzburg-Landau theory with Eq.(\ref{1}) \cite{sho};
in this case there are only the electromagnetic (normal state) and the
Higgs (superconducting state) phases and there is a critical value for an
external magnetic field at which the order parameter changes
discontinously from $\phi_o$ to zero and the superconductor returns to the
normal state (no Meissner effect) with a first-order phase transition.

When $\sqrt{2} \kappa > 1$ or $m_H > m_{em}$, 
one has type II superconductors, in 
which a third phase (vortices of magnetic field inside the material in the
superconducting state) is present and in which $< \phi (x) > =\phi_o(x)$
[i.e. $\phi_o$ is spatially varying]. To see the possibility of the
occurence of these vortices in the simulation with the Lagrangian density of
Eq.(\ref{1}) with the description involving only Dirac's observables, 
let us consider the physical Lagrangian density of Eq.(\ref{26}) 
in the Higgs phase in absence of fermions, for vanishing electric field
$\vec \pi (x)=\vec E(x)=0$ and in the static case $\partial^o{\vec A}^{'}(x)
=\partial^oH(x)=0$; moreover, let us suppose to have cylindrical symmetry,
$A^{'}_3(\vec x)=0$ and $\partial_3{\vec A}^{'}(\vec x)=0$ [so that
$B_1(\vec x)=B_2(\vec x)=0$, $B_3(\vec x)=-F_{12}(\vec x)$] and $\partial_3
H(\vec x)=0$. Let us also take $\phi_o=1$, $e={1\over 2}$, $\lambda
={1\over 8}$, so that $m_{em}=m_H=1/\sqrt{2}$ 
[this is the critical value
separating type I from type II superconductors]. Then the physical Lagrangian 
density (\ref{26}) reduces to [we follow Ref.\cite{tau} and
use the redefinition $H^{'}=H/\sqrt{2}$]

\begin{eqnarray}
{\cal L}^{"}(x^1,x^2)&=&-{1\over 2}F^2_{12}-{1\over 4}(1+{1\over \sqrt{2}}H)^2
(A^{{'}2}_1+A^{{'}2}_2)-{1\over 2}[(\partial_1H)^2+(\partial_2H)^2]-\nonumber \\
&-&{1\over 4}H^2(1+{1\over {2\sqrt{2}}}H)^2=\nonumber \\
&=&-{1\over 2}F^2_{12}-{1\over 4}(1+H^{'})^2(A^{{'}2}_1+
A^{{'}2}_2)-(\partial_1H^{'})^2-(\partial_2H^{'})^2-{1\over 8}
[(1+H^{'})^2-1]^2=\nonumber \\
&=&-{1\over 2}\lbrace (1+H^{'})^2[{1\over 2}(A^{{'}2}_1+A^{{'}2}_2)+
2(\partial_1ln(1+H^{'}))^2+2(\partial_2ln(1+H^{'}))^2-\nonumber \\
&-&\partial_1A^{'}_2+\partial_2A^{'}_1]+
[F_{12}+{1\over 2}((1+H^{'})^2-1)]^2+F_{12}\rbrace =\nonumber \\
&=&-{1\over 2}\lbrace (1+H^{'})^2[({{A^{'}_1}\over {\sqrt{2}}}-\sqrt{2}
\partial_2ln(1+H^{'}))^2+({{A^{'}_2}\over {\sqrt{2}}}+\sqrt{2}
\partial_1ln(1+H^{'}))^2]+\nonumber \\
&+&[F_{12}+{1\over 2}((1+H^{'})^2-1)]^2+
F_{12}-\partial_1[(1+H^{'})^2A^{'}_2]+\partial_2[(1+H^{'})^2A^{'}_1]\rbrace .
\label{36}
\end{eqnarray}

Therefore, modulo surface terms,the static 2-dimensional action
$S^{"}=-\int d^2x\, {\cal L}^{"}(x^1,x^2)$ is positive definite except for
the term $-{1\over 2}\int d^2x\, F_{12}$ . We get $-S^{"}=+{1\over 2}\int d^2x
F_{12}$ [the lower bound of Bogomol'nyi\cite{bog}; 
the conditions for having finite
action are $|\phi |=1+H^{'}{\rightarrow}_{r\rightarrow \infty} 1$, $e^{ie\theta}
D^{(A)}_{\mu}\phi =\partial_{\mu}H^{'}-ie(1+H^{'})A^{'}_{\mu} {\rightarrow}
_{r\rightarrow \infty} 0$] if

\begin{eqnarray}
&&A^{'}_1=2\partial_2 ln(1+H^{'}),\quad\quad A^{'}_2=-2\partial_1 ln(1+H^{'}),
\quad\quad \Rightarrow F_{12}=2\triangle ln(1+H^{'}),\nonumber \\
&&F_{12}=-{1\over 2}[(1+H^{'})^2-1],\nonumber \\
&&{\Downarrow}\nonumber \\
&&\triangle ln(1+H^{'})=-{1\over 4}[(1+H^{'})^2-1].
\label{37}
\end{eqnarray}

This is the form of the equations for the Nielsen-Olesen
vortices\cite{no} in terms of
Dirac observables when $\lambda ={1\over 8}$. For the vortex solution
[$|\phi (\vec x)|{\rightarrow}_{r\rightarrow \infty} 1$, $|{\vec A}^{'}(\vec x)|
{\rightarrow}_{r\rightarrow \infty}{1\over {er}}+{c\over e}\sqrt{ {{\pi}\over
{2er}} }e^{-er}$], one has
$S^{"}=-\pi n$ with $n={1\over {2\pi}}\int d^2x F_{12}$, a topological
invariant measuring the order of vanishing of $|\phi |=1+H$ in a discrete
set of points ${\vec x}_k$ around which $1+H(\vec x)={| \vec x-{\vec x}_k|}
^{n_k}+\ldots$ with $n=\sum_kn_k$ (see Ref.\cite{tau}). 
One recovers the electromagnetic phase [$m_{em}+|e|H=0$] in these points, where
the phase $\theta$ [satisfying $\theta (\vec x)=\theta (|\vec x|,\varphi )=
\theta (|\vec x|,\varphi +2\pi n)$] and the $\theta$-dependent gauge
transformation are singular.

\section
{The reformulation on spacelike hypersurfaces}

Both the phases are not described in a Lorentz-covariant way. Te remedy it, 
let us reformulate the Lagrangian density of Eq.(\ref{1}), in absence of
fermions for the sake of simplicity, on a family of
spacelike hypersurfaces foliating the Minkowski spacetime, along the
lines of Refs.\cite{lus,lus2,dir1}. We skip all the details of the construction,
which is fully explained in Ref.\cite{lus2}, and only sketch the starting
point and the final results.

If $z^{\mu}(\tau ,\vec \sigma )$ are the Minkowski coordinates of the
points of the spacelike hypersurface (each leaf of the foliation is identified
by the value of a scalar parameter $\tau$), whose curvilinear coordinates are
$\vec \sigma$, $g_{AB}(\tau ,\vec \sigma )={{\partial z^{\mu}(\tau ,\vec 
\sigma )}\over {\partial \sigma^A}}\eta_{\mu\nu}{{\partial z^{\nu}(\tau ,\vec 
\sigma )}\over {\partial \sigma^B}} =z^{\mu}_A(\tau ,\vec \sigma )\eta_{\mu\nu}
z_B^{\nu}(\tau ,\vec \sigma )$ [$A=\tau ,\check r$; $\sigma^{\tau}=\tau$]
the metric tensor induced on the hypersurface and $A_A(\tau ,\vec \sigma )=
z^{\mu}_A(\tau ,\vec \sigma )A_{\mu}(z(\tau ,\vec \sigma ))$ and $\tilde \phi
(\tau ,\vec \sigma )=\phi (z(\tau ,\vec \sigma ))$ the
electromagnetic potential and the Higgs field respectively, Eq.(\ref{1}) is
replaced by

\begin{eqnarray}
{\tilde {\cal L}}(\tau ,\vec \sigma )&=& {1\over 2}\sqrt{g(\tau ,\vec \sigma )}
\nonumber \\
&&\lbrace g^{\tau\tau}[D^{(A)}_{\tau}\tilde \phi ]^{*}D^{(A)}_{\tau}\tilde 
\phi +g^{\tau \check r}([D^{(A)}_{\tau}\tilde \phi ]^{*}D^{(A)}_{\check r}
\tilde \phi +[D^{(A)}_{\check r}\tilde \phi ]^{*}D^{(A)}_{\tau}\tilde \phi )+
g^{\check r\check s}[D^{(A)}_{\check r}\tilde \phi ]^{*}D^{(A)}_{\check s}
\tilde \phi -\nonumber \\
&-&V(\tilde \phi )-{1\over 2}g^{AC}g^{BD}F_{AB}F_{CD}\rbrace (\tau ,\vec 
\sigma ).
\label{38}
\end{eqnarray}

The canonical momenta are $\rho_{\mu}(\tau ,\vec \sigma )=\partial {\tilde
{\cal L}}(\tau ,\vec \sigma )/\partial z^{\mu}_{\tau}(\tau ,\vec \sigma )$,
$\pi^{\tau}(\tau ,\vec \sigma )=\partial {\tilde {\cal L}}/\partial
\partial_{\tau}A_{\tau}(\tau ,\vec \sigma )=0$, $\pi^{\check r}(\tau ,\vec 
\sigma )=\partial {\tilde {\cal L}}(\tau ,\vec \sigma )/\partial \partial
_{\tau}A_{\check r}(\tau ,\vec \sigma )$, ${\tilde \pi}_{\phi}(\tau ,\vec 
\sigma )=\partial {\tilde {\cal L}}(\tau ,\vec \sigma )/\partial \partial
_{\tau}\tilde \phi (\tau ,\vec \sigma )$, ${\tilde \pi}_{\phi^{*}}
(\tau ,\vec \sigma )=\partial {\tilde {\cal L}}(\tau ,\vec \sigma )/\partial 
\partial_{\tau}{\tilde \phi}^{*}(\tau ,\vec \sigma )$: $\pi^{\tau}(\tau ,\vec 
\sigma )$ and $\pi^{\check r}(\tau ,\vec \sigma )$ are now Lorentz-scalars.
We find five primary constraints

\begin{eqnarray}
&&{\cal H}_{\mu}(\tau ,\vec \sigma )=\rho_{\mu}(\tau ,\vec \sigma )-l_{\mu}
(\tau ,\vec \sigma )\Theta^{\tau\tau}(\tau ,\vec \sigma )-z_{{\check r}\mu}
(\tau ,\vec \sigma )\Theta^{\tau \check r}(\tau ,\vec \sigma )\approx 0
\nonumber \\
&&\pi^{\tau}(\tau ,\vec \sigma )\approx 0,
\label{39}
\end{eqnarray}

\noindent where $l_{\mu}(\tau ,\vec \sigma )$ is the normal to the
hypersurface, built only in terms of its tangent vectors $z^{\mu}_{\check r}
(\tau ,\vec \sigma )$. Since the canonical Hamiltonian vanishes, the Dirac
Hamiltonian, combination of the primary constraints, implies only the
secondary Lorentz-scalar constraint (Gauss law) $\Gamma (\tau ,\vec \sigma )=
-\partial^{\check r}\pi^{\check r}(\tau ,\vec \sigma )-ie({\tilde \pi}
_{\phi}\tilde \phi -{\tilde \pi}_{\phi^{*}}{\tilde \phi}^{*})(\tau ,\vec 
\sigma )\approx 0$. All the constraints are first class.

For the main stratum of field configurations with total timelike momentum,
$P^2 > 0$, we can reduce the theory to the Wigner hyperplanes orthogonal
to $P^{\mu}$. After the reduction, $\vec A(\tau ,\vec \sigma )$ and $\vec
\pi (\tau ,\vec \sigma )$ become Wigner spin-1 3-vectors and the pairs
$z^{\mu}(\tau ,\vec \sigma ), \rho_{\mu}(\tau ,\vec \sigma )$ are reduced
to a point ${\tilde x}^{\mu}_s(\tau ), p^{\mu}_s\approx P^{\mu}$, which
are the canonical coordinates of the center of mass of the configuration
of fields [${\tilde x}^{\mu}_s$ is not a four-vector] and, if we denote
$\epsilon_s=\pm \sqrt{p_s^2}$ the invariant mass of the system, we are
left only with the constraints

\begin{eqnarray}
{\cal H}(\tau )&=&\epsilon_s-\int d^3\sigma \lbrace {1\over 2}[{\vec \pi}^2
(\tau ,\vec \sigma )+{\vec B}^2(\tau ,\vec \sigma )]+{\tilde \pi}_{\phi^{*}}
(\tau ,\vec \sigma ){\tilde \pi}_{\phi}(\tau ,\vec \sigma )+\nonumber \\
&+&[(\vec \partial +ie\vec A(\tau ,\vec \sigma )){\tilde \phi}^{*}(\tau ,\vec 
\sigma )]\cdot (\vec \partial -ie\vec A(\tau ,\vec \sigma )){\tilde \phi} 
(\tau ,\vec \sigma )+\lambda ({\tilde \phi}^{*}(\tau ,\vec \sigma )\tilde \phi 
(\tau ,\vec \sigma )-\phi_o^2)^2\rbrace \approx 0\nonumber \\
{\vec {\cal H}}(\tau )&=&\int d^3\sigma \lbrace \vec \pi (\tau ,\vec \sigma )
\times \vec B(\tau ,\vec \sigma )+\nonumber \\
&+&{\tilde \pi}_{\phi}(\tau ,\vec \sigma )(\vec \partial -ie\vec A(\tau ,\vec 
\sigma ))\tilde \phi (\tau ,\vec \sigma )+{\tilde \pi}_{\phi^{*}}(\tau ,\vec 
\sigma )(\vec \partial +ie\vec A(\tau ,\vec \sigma )){\tilde \phi}^{*}(\tau ,
\vec \sigma )\rbrace \approx 0\nonumber \\
\pi^{\tau}(\tau ,\vec \sigma )&\approx& 0\nonumber \\
\Gamma (\tau ,\vec \sigma )&\approx& 0.
\label{40}
\end{eqnarray}

The constraints 
${\vec {\cal H}}\approx 0$ say that the hyperplane defines an intrinsic rest
frame for the system of fields; its gauge-fixings would force the center of
mass of the system defined inside the hyperplane to coincide with $x^{\mu}_s$
[the center of mass defined from outside the hyperplane, taking into
account its embedding in Minkowski spacetime]. This is the covariant 
rest-frame instant form of the dynamics\cite{lus2}.

We see that the reduction to either the electromagnetic or the Higgs
phase may be done as before, but now in a Lorentz invariant way.

In Eq.(\ref{38}) the configuration variables are $z^{\mu}(\tau ,\vec \sigma )$,
$A_A(\tau ,\vec \sigma )$ and $\tilde \phi (\tau ,\vec \sigma )$. As shown
in Appendix C of Ref.\cite{lus2}, instead of the gauge potentials $A_{\tau}
(\tau ,\vec \sigma )$, $A_{\check r}(\tau ,\vec \sigma )$ one can use
$A_l(\tau ,\vec \sigma )$, $A_{\check r}(\tau ,\vec \sigma )$ with
$A_{\mu}(z(\tau ,\vec \sigma ))=z^A_{\mu}(\tau ,\vec \sigma )A_A
(\tau ,\vec \sigma )=l_{\mu}(\tau ,\vec \sigma )A_l(\tau ,\vec \sigma )+
z^{\check r}_{\mu}(\tau ,\vec \sigma )A_{\check r}(\tau ,\vec \sigma )$
[here $Z^A_{\mu}(\tau ,\vec \sigma )$ are the vierbeins inverse of
$z_A^{\mu}(\tau ,\vec \sigma )$]. In this way Eq.(\ref{38}) is replaced by

\begin{eqnarray}
{\tilde {\cal L}}(\tau ,\vec \sigma )&=& {1\over 2}{{\sqrt{\gamma (\tau ,\vec 
\sigma )}}\over {N(\tau ,\vec \sigma )}}
\lbrace [D^{(A)}_{\tau}\tilde \phi ]^{*}D^{(A)}_{\tau}\tilde 
\phi -N^{\check r}([D^{(A)}_{\tau}\tilde \phi ]^{*}D^{(A)}_{\check r}
\tilde \phi +[D^{(A)}_{\check r}\tilde \phi ]^{*}D^{(A)}_{\tau}\tilde \phi )+
\nonumber \\
&+&(N^2\gamma^{\check r\check s}+N^{\check r}N^{\check s})
[D^{(A)}_{\check r}\tilde \phi ]^{*}D^{(A)}_{\check s}
\tilde \phi -N(\tau ,\vec \sigma )V(\tilde \phi )-\nonumber \\
&-&\sqrt{\gamma (\tau ,\vec \sigma )}[{1\over {2N}}\gamma^{\check r\check s}
(\partial_{\tau}A_{\check r}-{\cal L}_{\vec N}A_{\check r}-\partial_{\check r}
(NA_l))(\partial_{\tau}A_{\check s}-{\cal L}_{\vec N}A_{\check s}-\partial
_{\check s}(NA_l))+\nonumber \\
&+&{N\over 4}\gamma^{\check r\check s}\gamma^{\check u\check v}F_{\check r
\check u}F_{\check s\check v}](\tau ,\vec \sigma ) \rbrace,
\label{n1}
\end{eqnarray}

\noindent where $N(\tau ,\vec \sigma )=\sqrt{g(\tau ,\vec \sigma )/\gamma
(\tau ,\vec \sigma )}$ [$\gamma =-det\, |\, g_{\check r\check s}\, |$],
$N^{\check r}(\tau ,\vec \sigma )=g_{\tau \check s}(\tau ,\vec \sigma )
\gamma^{\check s\check r}(\tau ,\vec \sigma )$ [$\gamma^{\check r\check u}
g_{\check u\check s}=\delta^{\check r}_{\check s}$], are the lapse and shift
functions; one has $A_{\tau}(\tau ,\vec \sigma )=N(\tau ,\vec \sigma )A_l
(\tau ,\vec \sigma )+N^{\check r}(\tau ,\vec \sigma )A_{\check r}(\tau ,\vec 
\sigma )$ and $z^{\mu}_{\tau}(\tau ,\vec \sigma )=N(\tau ,\vec \sigma )
l^{\mu}(\tau ,\vec \sigma )+N^{\check r}(\tau ,\vec \sigma )z^{\mu}_{\check r}
(\tau ,\vec \sigma )$. Eq.(\ref{n1}) leads again to Eqs.(\ref{40}) [with
$\pi^l(\tau ,\vec \sigma )=N(\tau ,\vec \sigma )\pi^{\tau}(\tau ,\vec \sigma )
\approx 0$], since both $A_{\tau}(\tau ,\vec \sigma )$ and $A_l(\tau ,\vec 
\sigma )$ are gauge variables.

Instead the reformulation on spacelike hypersurfaces of the standard massive
vector field is

\begin{eqnarray}
{\tilde {\cal L}}^{'}(\tau ,\vec \sigma )&=&\sqrt{g(\tau ,\vec \sigma )}
\lbrace -{1\over 4}g^{AC}g^{BD}F_{AB}F_{CD}+{1\over  2}M^2g^{AB}A_AA_B
\rbrace (\tau ,\vec \sigma )=\nonumber \\
&=&-\sqrt{\gamma (\tau ,\vec \sigma )}\lbrace
{1\over {2N}}\gamma^{\check r\check s}
(\partial_{\tau}A_{\check r}-{\cal L}_{\vec N}A_{\check r}-\partial_{\check r}
(NA_l))(\partial_{\tau}A_{\check s}-{\cal L}_{\vec N}A_{\check s}-\partial
_{\check s}(NA_l))+\nonumber \\
&+&{N\over 4}\gamma^{\check r\check s}\gamma^{\check u\check v}F_{\check r
\check u}F_{\check s\check v}](\tau ,\vec \sigma ) \rbrace +\nonumber \\
&+&{1\over 2}M^2\sqrt{\gamma (\tau ,\vec \sigma )}N(\tau ,\vec \sigma )
\lbrace A^2_l(\tau ,\vec \sigma )+\gamma^{\check r\check s}(\tau ,\vec \sigma )
A_{\check r}(\tau ,\vec \sigma )A_{\check s}(\tau ,\vec \sigma )\rbrace .
\label{n2}
\end{eqnarray}

The final Lorentz-invariant constraints for $P^2 > 0$ on the hyperplane 
orthogonal to the total momentum, after the elination of $A_l, \pi^l$, are

\begin{eqnarray}
{\cal H}(\tau )&=&\epsilon_s-\int d^3\sigma \lbrace {1\over 2}[{\vec \pi}^2
(\tau ,\vec \sigma )+{\vec B}^2(\tau ,\vec \sigma )]+{1\over 2}M^2{\vec A}^2
(\tau ,\vec \sigma )+{{(\vec \partial \cdot \vec \pi (\tau ,\vec \sigma ))^2}
\over {2M^2}}\rbrace \approx 0\nonumber \\
{\vec {\cal H}}(\tau )&=&\int d^3\sigma \, \vec \pi (\tau ,\vec \sigma )\times
\vec B(\tau ,\vec \sigma )\, \approx 0.
\label{n3}
\end{eqnarray}

Now, on spacelike hypersurfaces there is the possibility to define in a
covariant way the Lagrangian density (\ref{m2}), which is replaced by

\begin{eqnarray}
{\tilde {\cal L}}^{"}(\tau ,\vec \sigma )&=&
-\sqrt{\gamma (\tau ,\vec \sigma )}\lbrace
{1\over {2N}}\gamma^{\check r\check s}
(\partial_{\tau}A_{\check r}-{\cal L}_{\vec N}A_{\check r}-\partial_{\check r}
(NA_l))(\partial_{\tau}A_{\check s}-{\cal L}_{\vec N}A_{\check s}-\partial
_{\check s}(NA_l))+\nonumber \\
&+&{N\over 4}\gamma^{\check r\check s}\gamma^{\check u\check v}F_{\check r
\check u}F_{\check s\check v}](\tau ,\vec \sigma ) \rbrace +\nonumber \\
&+&{1\over 2}\sqrt{\gamma (\tau ,\vec \sigma )}N(\tau ,\vec \sigma )
\lbrace M^2A^2_l-\gamma^{\check r\check s}\partial_{\check r}A_l\partial_{\check 
s}A_l+M^2\gamma^{\check r\check s}
A_{\check r}A_{\check s}\rbrace (\tau ,\vec \sigma ) .
\label{n4}
\end{eqnarray}

The final reduced Lorentz-invariant constraint on the hyperplane orthogonal
to the momentum are

\begin{eqnarray}
{\cal H}(\tau )&=&\epsilon_s-\int d^3\sigma \lbrace {1\over 2}[{\vec \pi}^2
(\tau ,\vec \sigma )+{\vec B}^2(\tau ,\vec \sigma )]+{1\over 2}M^2{\vec A}^2
(\tau ,\vec \sigma )+\nonumber \\
&+&{1\over 2}\vec \partial \cdot \vec \pi (\tau ,\vec \sigma ){1\over
{\triangle +M^2}}\vec \partial \cdot \vec \pi (\tau ,\vec \sigma )
\rbrace \approx 0\nonumber \\
{\vec {\cal H}}(\tau )&=&\int d^3\sigma \, \vec \pi (\tau ,\vec \sigma )\times
\vec B(\tau ,\vec \sigma )\, \approx 0.
\label{n5}
\end{eqnarray}

\section 
{Comments}

Let us make some final comments:

i) The same ambiguity in solving the Gauss law constraint, which originates
the two phases, is consistently present in the covariant R-gauge-fixing
\cite{rgau}

\begin{equation}
\partial^{\mu}A_{\mu}(x)+{1\over {\xi}} \theta (x)\approx 0
\label{41}
\end{equation}

\noindent used in the covariant-gauge approach to renormalization (to
remedy the nonrenormalizability of the unitary gauge) and in the
evaluation of radiative corrections with the associated Feynman rules
(see for instance Ref.\cite{cl}). Therefore, in these procedures one is mixing 
the two phases except in the final $\xi \rightarrow \infty$ limit to reach the
unitary gauge. Moreover, in the perturbative calculations one cannot see the
nonanalyticity in the coupling constant e (or in $m_{em}+|e|H(x)$) of the
electric phenomena in the Higgs phase.

ii) In Eq.(\ref{26}), the residual real scalar Higgs field $H(x)$ is actually 
coupled only to $|e|/m_{em}$; now this quantity appear in the mass term
of the vector gauge field and one is tempted to say that $H(x)$ is charged
but not minimally coupled to the electromagnetic field. To understand what
is going on we must study the conserved charges associated to the Gauss
law in both phases. This is not trivial due to the 
fact that in the broken gauge symmetry Higgs phase the electric and magnetic
fields decay at space infinity with Yukawa tails due to the mass $m_{em}$.
Therefore, the Gauss theorem breaks down: the electric charge in the Higgs
phase is a Noether constant of motion (first Noether theorem) but one cannot
measure it by means of the electric flux at space infinity (as in the case
of exact, not broken, local gauge symmetry; second Noether theorem\cite{lus3}).
This fact may be taken as a gauge-invariant signal of gauge symmetry breaking,
rather than the non-gauge-invariant quantum statement $< \phi > =\phi_o$
(see Ref.\cite{man} for a criticism of this criterion).

The Euler-Lagrange equations associated with the Lagrangian density
of Eq.(\ref{1}) are 

\begin{eqnarray}
L^{\mu}&=&{{\partial {\cal L}}\over {\partial A_{\mu}}}-\partial_{\nu}
{{\partial {\cal L}}\over {\partial \partial_{\nu}A_{\mu}}}=\partial_{\nu}
F^{\nu\mu}+eJ^{\mu}{\buildrel \circ \over =} 0\nonumber \\
&J&^{\mu}=\bar \psi \gamma^{\mu}\psi+i\phi^{*}[(\partial^{\mu}-ieA^{\mu})-
\stackrel{\longleftarrow}{(\partial^{\mu}+ieA^{\mu})}]\phi \nonumber \\
L_{\psi}&=&{{\partial {\cal L}}\over {\partial \psi}}-\partial_{\mu}
{{\partial {\cal L}}\over {\partial \partial_{\mu}\psi}}=-\bar \psi
[\stackrel{\longleftarrow}{(i\partial_{\mu}-eA_{\mu})}\gamma^{\mu}+
m]{\buildrel \circ \over =} 0\nonumber \\
L_{\bar \psi}&=&{{\partial {\cal L}}\over {\partial \bar \psi}}-\partial_{\mu}
{{\partial {\cal L}}\over {\partial \partial_{\mu}\bar \psi}}=[\gamma^{\mu}
(i\partial_{\mu}+eA_{\mu})-m]\psi {\buildrel \circ \over =} 0\nonumber \\
L_{\phi}&=&{{\partial {\cal L}}\over {\partial \phi}}-\partial_{\mu}
{{\partial {\cal L}}\over {\partial \partial_{\mu}\phi}}=-
{[D^{(A)\mu}D^{(A)}_{\mu}\phi ]}^{*}-{{\partial V(\phi )}\over
{\partial \phi}} {\buildrel \circ \over =} 0\nonumber \\
L_{\phi^{*}}&=&{{\partial {\cal L}}\over {\partial \phi^{*}}}-\partial_{\mu}
{{\partial {\cal L}}\over {\partial \partial_{\mu}\phi^{*}}}=-
D^{(A)\mu}D^{(A)}_{\mu}\phi -{{\partial V(\phi )}\over {\partial \phi^{*}}}
{\buildrel \circ \over =} 0.
\label{42}
\end{eqnarray}

Let us note that in presence of external electromagnetic fields [so that
$A_{\mu}\mapsto A_{\mu}+A_{ext,\mu}$]
the Euler-Lagrange equations of the Higgs field are solved by requiring
\cite{sho1}

\begin{eqnarray}
&&D^{(A+A_{ext})}_{\mu}\phi {\buildrel \circ \over =} 0\nonumber \\
&&{{\partial V(\phi )}\over {\partial \phi}} {\buildrel \circ \over =} 0.
\label{43}
\end{eqnarray}

\noindent While the second  equation has the two solutions $\phi =0$ and
$\phi =\phi_o$,
from the first equation we get $0 {\buildrel \circ \over =}
[D^{(A+A_{ext})}_{\mu},D^{(A+A_{ext})}_{\nu}]\phi =-ie[F_{\mu\nu}+
F_{ext,\mu\nu}]\phi$. Therefore, we get either $\phi =0$ and $F+F_{ext}\not= 0$
(the electromagnetic phase) or $\phi =\phi_o$ and the Meissner effect
$F+F_{ext}=0$ (the Higgs phase).

The gauge invariance of ${\cal L}(x)$ under the infinitesimal
gauge transformations $\delta A_{\mu}=-{1\over e}\partial_{\mu}\alpha$,
$\delta \psi =-i\alpha \psi$, $\delta \bar \psi =i\bar \psi \alpha$,
$\delta \phi =-i\alpha \phi$, $\delta \phi^{*}=i\alpha \phi^{*}$,
produces the Noether
identities

\begin{eqnarray}
0\equiv \delta {\cal L}&=&{{\partial {\cal L}}\over {\partial A_{\mu}}}
\delta A_{\mu}+{{\partial {\cal L}}\over {\partial \partial_{\nu}A_{\mu}}}
\delta \partial_{\nu}A_{\mu}+\delta \bar \psi {{\partial {\cal L}}\over
{\partial \bar \psi}}+\delta \partial_{\mu}\bar \psi {{\partial {\cal L}}
\over {\partial \partial_{\mu}\bar \psi}}+\nonumber \\
&+&\delta \psi {{\partial {\cal L}}\over
{\partial \psi}}+\delta \partial_{\mu}\psi {{\partial {\cal L}}
\over {\partial \partial_{\mu}\psi}}+{{\partial {\cal L}}\over {\partial \phi}}
\delta \phi +{{\partial {\cal L}}\over {\partial \partial_{\mu}\phi}}
\delta \partial_{\mu}\phi +\nonumber \\
&+&{{\partial {\cal L}}\over {\partial \phi^{*}}}
\delta \phi^{*} +{{\partial {\cal L}}\over {\partial \partial_{\mu}\phi^{*}}}
\delta \partial_{\mu}\phi^{*}=\nonumber \\
&=&L^{\mu}\delta A_{\mu}+\delta \bar \psi L_{\bar \psi}-L_{\psi}\delta \psi +
\delta \phi^{*}L_{\phi^{*}}+L_{\phi}\delta \phi +\partial_{\mu}G^{\mu}
\nonumber \\
&&{}\nonumber \\
G^{\mu}&=&\alpha G^{\mu}_{1}+\partial_{\nu}\alpha G^{\mu\nu}_{o}=\nonumber \\
&=&-F^{\mu\nu}\delta A_{\nu}-{i\over 2}[\delta \bar \psi \gamma^{\mu}\psi -
\bar \psi \gamma^{\mu}\delta \psi ]+{[D^{(A)\mu}\phi ]}^{*}\delta \phi
+\delta \phi^{*} D^{(A)\mu}\phi \nonumber \\
&&\Downarrow \nonumber \\
G^{\mu\nu}_o&=&{1\over e}F^{\mu\nu}\nonumber \\
G^{\mu}_1&=&\bar \psi \gamma^{\mu}\psi +i(\phi^{*}D^{(A)\mu}\phi-
{[D^{(A)\mu}\phi ]}^{*}\phi)=J^{\mu}\nonumber \\
&&{}\nonumber \\
\partial_{\mu}G^{\mu}&=&\partial_{\mu}\partial_{\nu}\alpha G^{\mu\nu}_o+
\partial_{\mu}\alpha [\partial_{\nu}G^{\nu\mu}_o+G^{\mu}_1]+\alpha \partial
_{\mu}G^{\mu}_1\equiv \nonumber \\
&\equiv& -L^{\mu}\delta A_{\mu}+L_{\psi}\delta \psi -\delta \bar \psi 
L_{\bar \psi}-L_{\phi}\delta \phi -\delta \phi^{*}L_{\phi^{*}}
{\buildrel \circ \over =} 0.
\label{44}
\end{eqnarray}

The last line implies the Noether identities [$(\mu\nu )$ and $[\mu\nu ]$ mean
symmetrization and antisymmetrization respectively]

\begin{eqnarray}
&&G^{(\mu\nu )}_o\equiv 0\nonumber \\
&&\partial_{\nu}G^{\nu\mu}_o\equiv -G^{\mu}_1+{1\over e}L^{\mu}={1\over e}
L^{\mu}-\bar \psi \gamma^{\mu}\psi -i(\phi^{*}D^{(A)\mu}\phi -
[D^{(A)\mu}\phi ]^{*}\phi )\nonumber \\
&&\partial_{\mu}G^{\mu}_1\equiv -i(L_{\psi}\psi +\bar \psi L_{\bar \psi})+i
(L_{\phi}\phi -\phi^{*}L_{\phi^{*}}){\buildrel \circ \over =} 0
\label{45}
\end{eqnarray}

\noindent and, from the last two lines of these equations, the contracted
Bianchi identities

\begin{equation}
\partial_{\mu}L^{\mu}+ie(L_{\psi}\psi +\bar \psi L_{\bar \psi}+
\phi^{*}L_{\phi^{*}}-L_{\phi}\phi )\equiv 0.
\label{46}
\end{equation}

The following subset of Noether identities reproduces the Hamiltonian
constraints

\begin{eqnarray}
&&\pi^o=-eG^{(oo)}_o\equiv 0\nonumber \\
&&0\equiv \partial^o \pi^o\equiv \partial^k\pi^k-eJ^o-L^o=-\Gamma -L^{o}
{\buildrel \circ \over =}-\Gamma.
\label{47}
\end{eqnarray}

The strong improper conservation law \cite{lus3} $\partial_{\mu}V^{\mu}\equiv 
0$, implied by Eqs.(\ref{45}), identifies the strong improper conserved current
(strong continuity equation)

\begin{eqnarray}
V^{\mu}&=&-\partial_{\nu}G^{\nu\mu}_o=-{1\over e}\partial_{\nu}F^{\nu\mu}=
\partial_{\nu}U^{[\mu\nu ]}{\buildrel \circ \over =} J^{\mu}= \nonumber \\
&=&\bar \psi \gamma^{\mu}\psi +i(\phi^{*}D^{(A)\mu}\phi-
[D^{(A)\mu}\phi]^{*}\phi )=G^{\mu}_1=j_F^{\mu}+j_{KG}^{\mu},
\label{48}
\end{eqnarray}

\noindent with the superpotential $U^{[\mu\nu ]}={1\over e}F^{\mu\nu}$. In 
the last line, $j^{\mu}_f=\bar \psi \gamma^{\mu}\psi$ and $j_{KG}^{\mu}=i
(\phi^{*}D^{(A)\mu}\phi -[D^{(A)\mu}\phi]^{*}\phi )$
are the charge currents of the fermion field and of
the complex Klein-Gordon Higgs fields respectively.

The associated weak improper conservation law is $\partial_{\mu}G^{\mu}_1
{\buildrel \circ \over =} 0$ [it is obtained by using the second line of 
Eqs.(\ref{45})]. If Q is the weak improper conserved Noether charge and
$Q^{(V)}$ the strong improper conserved one, we get [its meaning is
equivalent to $\int d^3x \Gamma (\vec x,x^o){\buildrel \circ \over =} 0$]

\begin{eqnarray}
Q&=& \int d^3x G^o_1(\vec x,x^o)=\int d^3x J^o(\vec x,x^o)=\nonumber \\
&=&\int d^3x[\bar \psi (\vec x,x^o)\gamma^o
\psi (\vec x,x^o)-i(\pi_{\phi}(\vec x,x^o)\phi (\vec x,x^o)-\phi^{*}(\vec x,x^o)
\pi_{\phi^{*}}(\vec x,x^o))]=\nonumber \\
&=&\int d^3x [\psi^{\dagger}(\vec x,x^o)\psi (\vec x,x^o)-\pi_{\theta}
(\vec x,x^o)]=Q_F+Q_{\theta}{\buildrel \circ \over =}\nonumber \\
{\buildrel \circ \over =} Q^{(V)}&=& \int d^3x V^o(\vec x,x^o)=\int d^3x
\partial^kF^{ko}(\vec x,x^o)=\int d^3x
\vec \partial \cdot \vec \pi (\vec x,x^o)=\int d\vec \Sigma \cdot \vec 
E(\vec x,x^o),\nonumber \\
&&{}
\label{49}
\end{eqnarray}

\noindent where $Q_F$ and $Q_{\theta}$ are the electric charges (in units of e)
of the fermion fields and of the complex Higgs field.

In the electromagnetic phase, $Q{\buildrel \circ \over =} Q^{(V)}$ is the Gauss
theorem associated with the long-range electromagnetic interaction: 
the flux at space infinity of the electric field is equal to the
total electric charge of the fermions and of the charged complex Higgs
fields, dressed with their Coulomb clouds [$Q=\int d^3x [{\check {\bar 
\psi}}^{\dagger}{\check \psi}-i({\check \pi}_{\phi}{\check \phi}-{\check
\phi}^{*}{\check \pi}_{\phi^{*}})](\vec x,x^o)$], 
with the additional information that the Higgs electric charge is
carried by the phase $\theta (x)$. 

On the contrary, in the broken symmetry Higgs phase we get $Q^{(V)}=0$
when Eq.(\ref{49}) is integrated over all the 3-space, because the electric 
field decays exponentially at space infinity due to the generated 
electromagnetic mass $m_{em}$ (short-range interaction), 
so that the Gauss theorem breaks down in presence of
spontaneous symmetry breaking through the Higgs mechanism. 
The residual Higgs field $H(x)$ turns out to be neutral, being instead coupled 
to the ratio $|e|/m_{em}$ and the
electromagnetic mass is replaced by the effective mass
$m_{em}(1+{{|e|}\over {m_{em}}}H)$ [one could also say that the last term of
Eq.(\ref{25}) describes an effective mass $m_H(1+{{|e|}\over {2m_{em}}}H)$ for
H itself]. When we integrate over all 3-space,
Eq.(\ref{49}) can be written as

\begin{eqnarray}
&&Q={\hat Q}_F+Q_{\theta} {\buildrel \circ \over =} 0\nonumber \\
&&Q_{\theta} {\buildrel \circ \over =} -{\hat Q}_F=-\int d^3x [{\hat {\bar
\psi}}^{\dagger}{\hat \psi}](\vec x,x^o),
\label{50}
\end{eqnarray}

\noindent and says that the charge $Q_{\theta}$ of the nonlinearly 
interacting would-be massless Goldstone boson 
$\theta (x)$ [which does not appear among the Dirac's observables, being eaten 
by the vector field, 
and which has the quantum numbers of the broken generator of U(1) at the
quantum level (see Ref.\cite{str} for the infrared singularity associated
with this unphysical massless would-be Goldstone boson)]
is opposite to the fermionic electric charge ${\hat Q}_F$, which, by itself,
is an ordinary conserved Noether charge due to the invariance of the
physical Lagrangian density of Eq.(\ref{26}) under global phase
transformations of the fermion fields: $\psi \mapsto e^{-i\alpha}\psi$ implies

\begin{equation}
{d\over {dx^o}}{\hat Q}_F{\buildrel \circ \over =} 0. 
\label{51}
\end{equation}

Eq.(\ref{50}) is
consistent with the fact that each fermion field is dressed with a Higgs
cloud of $\theta$-field which screens the fermion electric charge if looked 
from space infinity in the way of Eq.(\ref{49}) in the Higgs phase of the
original theory before the reduction to Dirac's observables; the absence of 
Gauss' theorem is also evident in the self-energy term in the physical 
Hamiltonian of Eq.(\ref{28}). When Eq.(\ref{49}) is
integrated over a finite domain ${\cal V}$, we have $Q_{\theta}{\buildrel
\circ \over =} -{\hat Q}_F+\int_{\cal V}d\vec \Sigma \cdot \vec E(\vec x,x^o)$.

iii) As a last remark, we note that the Lagrangian densities associated to 
Dirac's observables are in general nonlocal and nonpolynomial, so that the 
standard  regularization and renormalization prescriptions do not hold. In 
Refs.\cite{lus,lus2} it is shown that for every extended relativistic system 
(particles, strings, field configurations) in an irreducible timelike
poincar\'e representation one can define a classical unit of length
$\rho =\sqrt{-W^2}/P^2$ in terms of the Poincar\'e Casimirs from the
discussion of the center-of-mass problem ($\rho$ is a measure of the domain
in 3-space defined by the noncovariance of the center-of-mass coordinate
${\tilde x}^{\mu}_s$). This can, we hope, be the basis
for a ultraviolet cutoff for the quantization of 
theories formulated on spacelike
hypersurfaces (classical background of the Tomonaga-Schwinger approach).

\vfill\eject

\end{document}